\documentclass[10pt, aps, prd,
notitlepage, nofootinbib, longbibliography, showpacs
]{revtex4-1}

\usepackage{pstricks,pst-node,pst-plot,pst-3d,pst-3dplot,pst-text,pst-eps}
\usepackage{amsmath, amsthm, amssymb, bm}
\usepackage{graphicx}
\usepackage{subfigure}
\usepackage{hyperref} 

\begin{document}

\title{Electromagnetic semitransparent $\delta$-function plate:
Casimir interaction energy between parallel infinitesimally thin plates}

\author{Prachi Parashar} \email{prachi@nhn.ou.edu}
\author{Kimball A. Milton}
\email{milton@nhn.ou.edu} 
\homepage{http://www.nhn.ou.edu/~milton}
\affiliation{Homer L. Dodge Department of Physics and Astronomy,
University of Oklahoma, Norman, Oklahoma 73019, USA}

\author{K. V. Shajesh\footnote{Present address:
Department of Materials Science and Engineering,
Iowa State University of Science and Technology,
Ames, Iowa 50011, USA} }
\email{shajesh@andromeda.rutgers.edu}
\homepage{http://www.nhn.ou.edu/~shajesh}
\author{M. Schaden}
\email{mschaden@andromeda.rutgers.edu}
\homepage{http://www.ncas.rutgers.edu/martin-schaden}
\affiliation{Department of Physics, Rutgers, The State University of New Jersey, Newark, New Jersey 07102, USA} 

\date{\today}

\begin{abstract}

We derive boundary conditions for electromagnetic fields on a
$\delta$-function plate. The optical properties of such a plate
are shown to necessarily be anisotropic in that they only depend
on the transverse properties of the plate. We unambiguously obtain
the boundary conditions for a perfectly conducting $\delta$-function
plate in the limit of infinite dielectric response. We show that
a material does not ``optically vanish'' in the thin-plate limit.
The thin-plate limit of a plasma slab of thickness $d$ with plasma
frequency $\omega_p^2=\zeta_p/d$ reduces to a $\delta$-function
plate for frequencies ($\omega=i\zeta$) satisfying
$\zeta d \ll \sqrt{\zeta_p d} \ll 1$. We show  that the Casimir
interaction energy between two parallel perfectly conducting
$\delta$-function plates is the same as that for parallel perfectly
conducting slabs. Similarly, we show that the interaction energy
between an atom and a perfect electrically conducting
$\delta$-function plate is the usual Casimir-Polder energy, which is
verified by considering the thin-plate limit of dielectric slabs. The
``thick'' and ``thin'' boundary conditions considered by Bordag are
found to be identical in the sense that they lead to the same
electromagnetic fields.

\end{abstract}

\maketitle

\section{Introduction}

Idealized infinitesimally thin perfectly conducting surfaces are often
envisaged to decouple electromagnetically two regions in space. Boyer
in 1968 \cite{Boyer:1968uf} found that the Casimir energy of such a thin
perfectly conducting spherical shell contributes a radial outward pressure
on the surface of the shell. The repulsive nature of this force, unlike the
attractive Casimir force between parallel perfectly conducting plates
\cite{Casimir:1948dh}, has remained poorly understood, but general 
systematics are becoming clearer
\cite{Bender:1994zr,Abalo:2010ah,Abalo:2012jz}.
To investigate the physical nature of Boyer's result,
Barton in Ref.~\cite{Barton:2004sph,Barton:2004sp2,Barton:2004sp3}
used a hydrodynamic model developed in 
Refs.~\cite{Fetter:1973slg,Fetter:1974pag}
to study the Casimir energy of a spherical plasma shell
as a continuum model of interactions in C$_{60}$ and C$_{70}$ molecules.
Barton observed that an infinitesimally thin conducting surface
imposes non-trivial boundary conditions on the electromagnetic fields
and in Refs.~\cite{Barton:2005eps,Barton:2005fst}
considered ``a fluid model of an
infinitesimally thin plasma sheet to describe a single base plane
from graphite,'' in which boundary conditions on such a plate are 
derived by integrating across the plate.
These boundary conditions are broadly referred to as the plasma shell model.

A $\delta$-function as a model for an infinitesimally thin conducting surface
was first used in Refs.~\cite{Bordag:1985th,Bordag:1992cp}.
Reference~\cite{Robaschik:1994bc} proposed that  
perfect electrical conductors could satisfy two independent boundary 
conditions, dubbed ``thick'' and ``thin'' boundary 
conditions in Ref.~\cite{Bordag:2006psr}.
Bordag in Ref.~\cite{Bordag:2004smr} reported that
the Casimir-Polder force calculated with ``thin'' boundary conditions
is 13\% lower than that obtained with ``thick'' boundary conditions.
Variations of this calculation has been presented again in 
Refs.~\cite{Bordag:2006psr} and \cite{Bordag:2007spt}.
This has led to confusion in the understanding of
the boundary conditions on a $\delta$-function plate,
which eventually has also entered into discussions on graphene. We resolve 
the issue by showing that all the three boundary conditions:
Barton's plasma shell model, and Bordag's ``thick'' and ``thin''
boundary conditions, are identical.


In this paper we consider an idealized infinitesimally thin material whose
electric and magnetic properties are described by an electric permittivity
and a magnetic permeability written in terms of a $\delta$-function,
\begin{subequations}
\begin{eqnarray}
{\bm \varepsilon}(z) - {\bf 1} &=& {\bm \lambda}_e \delta(z), 
\label{epmu-def-e} \\
{\bm \mu}(z) - {\bf 1} &=& {\bm \lambda}_g \delta(z).
\end{eqnarray}%
\label{epmu-def}%
\end{subequations}
Following Refs.~\cite{Scandurra:1999sss,Bordag:1999bsf}
we call this a semitransparent $\delta$-function plate.
${\bm\lambda}_e$ and $\bm\lambda_g$ in this model have dimensions of length,
and are in general frequency dependent.
The $\delta$-function in Eq.~(\ref{epmu-def}) should be thought of as
a limit of a sequence of functions that are symmetric about the
plane $z=0$~\cite{Estrada:2007pt}.
The electric permittivity and magnetic permeability is assumed 
isotropic in the plane of the plate only. The distinction of 
parallel and perpendicular components will be made in reference 
to the $z$-direction chosen normal to the plane of the plate,
for example, ${\bf k}_\perp$ will represent components of ${\bf k}$
perpendicular to the normal.
The tensor structure of ${\bm\lambda}_e$ and $\bm\lambda_g$ is 
chosen to be diagonal in the $xyz$-coordinate system for simplicity.

In Sec.~\ref{dp-max} we study Maxwell's equations in the presence of
$\delta$-function plates described by Eqs.~(\ref{epmu-def}).
Employing a Gaussian surface integral and an Amperian loop integral
across the plate
 we find additional non-vanishing contributions in the
boundary conditions due to the $\delta$-functions in Eqs.~(\ref{epmu-def}). 
They imply anisotropic optical properties for 
a semitransparent $\delta$-function plate. These boundary conditions
are derived for the case of a $\delta$-function plate sandwiched between
two uniaxial materials and reduce to the conventional 
boundary conditions when ${\bm\lambda}_e=0$ and $\bm\lambda_g=0$, and are
consistent with those derived by Barton in Ref.~\cite{Barton:2005eps}
for a purely electric $\delta$-function plate.
The anisotropy in Barton's plasma sheet model
arises implicitly due to the requirement that any displacements
in the plasma are tangential to the plate.

In Sec.~\ref{fields-and-gf-sec}
magnetic and electric Green's functions, representing the two
independent modes, are constructed and the fields are defined in 
terms of the Green's dyadics. In Sec.~\ref{HEGreen}
the boundary conditions derived for the fields
are transcribed onto the magnetic and electric Green's functions.
We unambiguously derive solutions for these Green's functions,
which determine the 
reflection and transmission coefficients for a semitransparent 
$\delta$-function plate. Our solutions generalize Barton's by
allowing for non-trivial magnetic properties of a $\delta$-function plate.
The rate of change of energy density of the electromagnetic field
is shown to be balanced by the energy flux across the plate.
This statement of the conservation of energy is expressed
in terms of reflection and transmission coefficients of
a $\delta$-function plate. Both
a perfect electrically conducting $\delta$-function plate with vanishing
magnetic susceptibility, and a perfect magnetically conducting
$\delta$-function plate with vanishing electric response,
are perfect reflectors that do not transmit energy across the plate.
By contrast, a perfect
electrically and magnetically conducting
$\delta$-function plate is a perfect
transmitter of energy while introducing a phase shift of $180^\circ$ in
the transmitted fields, and would pass the test for invisibility.
A split coherent laser beam incident on such a plate from both sides
would extinguish itself.

In Sec.~\ref{bc-tp-sec} we derive boundary conditions across a
perfect electrically conducting $\delta$-function plate by taking 
the limit of infinite dielectric response.
We find that even though the tangential component of the electric field
vanishes on a $\delta$-function plate, the corresponding component of
electric displacement field does not, unambiguously specifying
the discontinuity in the normal component of the electric displacement field
across the $\delta$-function plate.
We clarify how a semi-infinite isotropic dielectric slab
and an (anisotropic) $\delta$-function plate lead to
the same boundary conditions and the same physics
in the limit of perfect conduction. 

Following Ref.~\cite{Prachi:2011ec},
we investigate the feasibility of physically
realizing a semitransparent $\delta$-function plate
in Sec.~\ref{physi-dp}.
In contrast to naive expectations we show that a material
does not necessarily optically vanish in the limit of zero thickness.
The thin-plate limit, in which
the optical properties of a dielectric slab of thickness $d$
with plasma frequency $\omega_p^2=\zeta_p/d$ may be approximated
by a $\delta$-function plate, is shown to be satisfied for frequencies
in the range $\zeta d \ll \sqrt{\zeta_p d} \ll 1$.
By modelling the charge carriers in a conducting slab as a Fermi gas
with Neumann boundary conditions we explore the conditions necessary
for the realization of a two-dimensional conducting sheet. In this model we
relate the optical properties of a $\delta$-function plate
with the plasma frequency of the two-dimensional plasma. 

In Sec.~\ref{Cas-tp} the interaction energy between two parallel
$\delta$-function plates is obtained from their optical properties.
We explicitly show that the interaction energy between parallel
perfect electrically conducting $\delta$-function plates is the 
Casimir energy between two ideal metallic slabs.
We further verify that the interaction energy
between parallel anisotropic slabs
in the limit of vanishing thickness (thin-plate limit) also
reproduces the interaction energy for parallel $\delta$-function plates.
In Sec.~\ref{CP-ThickThin} the interaction energy 
between an atom and a semitransparent $\delta$-function plate is calculated.
In the perfect conductor limit this interaction energy is just the
usual Casimir-Polder energy between an atom and an ideal metallic slab.
This is verified independently using the thin-plate limit.
We show that all of the three boundary conditions discussed in
the literature to understand the electrodynamics of a $\delta$-function 
plate---Barton's plasma shell model and Bordag's ``thick'' and 
``thin'' boundary conditions---are physically identical.
In the final section we conclude by discussing our results.

In Appendix~\ref{avpres-sec} we show that a function with a step 
discontinuity when evaluated at the point of discontinuity should 
be interpreted as the average of the function's left and right
limiting values.
In Appendices~\ref{G1-eva-sec} and \ref{lifE-aniso-sec} the details
of the calculations leading to the Casimir interaction energy between 
two semitransparent $\delta$-function plates and two anisotropic slabs
are presented.


\section{Infinitesimally thin $\delta$-function plates}
\label{dp-max}

We consider a semitransparent $\delta$-function plate described by 
Eqs.~(\ref{epmu-def}) and restrict our analysis to uniaxial materials
with the optical axis oriented normal to the plane of the plate,
chosen as the $\hat{\bf z}$ direction of a Cartesian coordinate system.
We thus consider materials whose optical properties are represented by
\begin{subequations}
\begin{eqnarray}
{\bm\lambda}_e &=& \lambda^\perp_e\, {\bf 1}_\perp 
+ \lambda^{||}_e \, \hat{\bf z}\, \hat{\bf z}, \label{mat-tp-e} \\
{\bm\lambda}_g &=& \lambda^\perp_g \, {\bf 1}_\perp 
+ \lambda^{||}_g \,\hat{\bf z} \,\hat{\bf z}, 
\end{eqnarray}%
\label{mat-tp}%
\end{subequations}
where $\lambda_{e,g}^\perp$ and $\lambda_{e,g}^{||}$,
in general are frequency dependent.
In Heaviside-Lorentz units
the monochromatic components proportional to $\exp(-i\omega t)$ of
Maxwell's equations in the absence of charges and currents are
\begin{subequations}
\begin{eqnarray}
{\bm \nabla} \times {\bf E} &=& i \omega {\bf B}, \\
-{\bm \nabla} \times {\bf H} &=& i \omega ({\bf D} + {\bf P}), 
\end{eqnarray}%
\label{ME-cross}%
\end{subequations}
which implies ${\bm \nabla} \cdot {\bf B}=0$, and
${\bm \nabla} \cdot ({\bf D}+ {\bf P})=0$,
where ${\bf P}$ is an external source of polarization.
We in the following neglect non-linear responses and assume that 
the fields ${\bf D}$ and ${\bf B}$ are linearly dependent on the 
electric and magnetic fields ${\bf E}$ and ${\bf H}$ as
\begin{subequations}
\begin{eqnarray}
{\bf D}({\bf x},\omega) 
&=& {\bm \varepsilon}({\bf x};\omega) \cdot {\bf E}({\bf x},\omega), \\
{\bf B}({\bf x},\omega) 
&=& {\bm \mu}({\bf x};\omega) \cdot {\bf H}({\bf x},\omega).
\end{eqnarray}%
\label{DB=emuEB}%
\end{subequations} 
Exploiting translational symmetry in the plane of the plate and
rotational symmetry about the normal $\hat{\bf z}$ direction,
we may consider a plane wave with wave-numbers 
$k_x$ and $k_y$ and choose $k_y=0$ without loss of generality.
We thus can write 
${\bm \nabla} = ik_\perp \hat{\bf x} + \hat{\bf z} \partial_z$,
with $k_\perp = k_x$.
Maxwell's equations in Eqs.~(\ref{ME-cross}) thus decouple into those for
two modes\footnote{%
In the context of wave-guides these modes refer to
the direction of propagation ($\hat{\bf x}$
in our case). TM (TE) modes have no magnetic (electric) field
in the direction of propagation, whereas
E(H)-modes have a electric (magnetic) field component
in the direction of propagation.
Note that the same can be thought of relative to the
direction normal to the plane ($\hat{\bf z}$).}:
the transverse magnetic mode (TM or E-mode) involves 
the field components $(E_1, H_2, E_3)$:
\begin{subequations}
\begin{eqnarray}
H_2(z) &=& - \frac{\omega}{k_\perp} D_3(z) - \frac{\omega}{k_\perp} P_3(z),
\label{TM-first-H2} \\
\frac{\partial}{\partial z} D_3(z)
&=& -ik_\perp D_1(z) -ik_\perp P_1(z) - \frac{\partial}{\partial z} P_3(z),
\label{TM-first-D3} \\
\frac{\partial}{\partial z} E_1(z)
&=& ik_\perp E_3(z) + i\omega B_2(z),
\label{TM-first-E1}
\end{eqnarray}%
\label{TM-first-order}%
\end{subequations}
and the transverse electric mode (TE or H-mode) involves
the field components $(H_1, E_2, H_3)$: 
\begin{subequations}
\begin{eqnarray}
E_2(z) &=& \frac{\omega}{k_\perp} B_3(z),
\label{TE-first-E2} \\
\frac{\partial}{\partial z} B_3(z)
&=& -ik_\perp B_1(z), 
\label{TE-first-B3} \\
\frac{\partial}{\partial z} H_1(z)
&=& ik_\perp H_3(z) - i\omega D_2(z) - i\omega P_2(z).
\end{eqnarray}%
\label{TE-first-order}%
\end{subequations}
The Maxwell equations in
Eqs.~(\ref{TM-first-order}) and (\ref{TE-first-order}),
which are in first order form, can be combined to yield
the second order differential equations
[with ${\bm\varepsilon} = 
\text{diag}(\varepsilon^\perp,\varepsilon^\perp,\varepsilon^{||})$
and ${\bm\mu}= \text{diag}(\mu^\perp,\mu^\perp,\mu^{||})$]
\begin{subequations}
\begin{eqnarray}
\left[ - \frac{\partial}{\partial z} \frac{1}{\varepsilon^\perp(z)}
\frac{\partial}{\partial z} + \frac{k_\perp^2}{\varepsilon^{||}(z)} 
-\omega^2 \mu^\perp(z) \right] H_2(z) 
&=&- i\omega \frac{\partial}{\partial z} \frac{P_1(z)}{\varepsilon^\perp(z)}
- \omega k_\perp \frac{P_3(z)}{\varepsilon^{||}(z)},
\label{2order-D3} \\
\left[ - \frac{\partial}{\partial z} \frac{1}{\mu^\perp(z)}
\frac{\partial}{\partial z} + \frac{k_\perp^2}{\mu^{||}(z)} 
-\omega^2 \varepsilon^\perp(z) \right] E_2(z) &=& \omega^2 P_2(z).
\label{2order-B3}%
\end{eqnarray}%
\label{2order-D3B3}%
\end{subequations}
The remaining field components can be expressed in terms of
$H_2(z)$ and $E_2(z)$:
$E_1(z)$ using Eqs.~(\ref{TM-first-H2}) and (\ref{TM-first-D3}),
$E_3(z)$ using Eq.~(\ref{TM-first-H2}),
$H_1(z)$ using Eqs.~(\ref{TE-first-E2}) and (\ref{TE-first-B3}),
and $H_3(z)$ using Eq.~(\ref{TE-first-E2}).


\subsection{Boundary conditions}
\label{bc-section}

Boundary conditions for electromagnetic fields due to the presence of
a single semitransparent $\delta$-function plate in vacuum,
or such a plate sandwiched between
two adjacent semi-infinite slabs, are derived by integrating
Maxwell's equations in Eqs.~(\ref{ME-cross}),
or more explicitly Eqs.~(\ref{TM-first-order}) and (\ref{TE-first-order}),
across the $\delta$-function plate positioned at $z=a$.
We require electric and magnetic fields
to be free of $\delta$-function type singularities,
\begin{equation}
\lim_{\delta\to 0}\int_{a-\delta}^{a+\delta}dz\, {\bf E}(z) =0, 
\qquad \text{and} \qquad
\lim_{\delta\to 0}\int_{a-\delta}^{a+\delta}dz\, {\bf H}(z) =0,
\label{sing-con}
\end{equation}
implying that the 
$\delta$-function singularities of ${\bm \varepsilon}$ and ${\bm \mu}$
in Eqs.~(\ref{epmu-def}) are completely contained in ${\bf D}$ and ${\bf B}$
as a consequence of Eqs.~(\ref{DB=emuEB}).
To illustrate this explicitly let us for the moment consider the
case of ${\bm\lambda}_g={\bf 0}$ and $\lambda_e^{||}=0$.
Then, for any given source distribution ${\bf P}$,
Eqs.~(\ref{TM-first-order}) can be combined to yield
\begin{equation}
\Big[ -\frac{\partial^2}{\partial z^2} + (k_\perp^2 - \omega^2)
+ (k_\perp^2 - \omega^2) \lambda_e^\perp \delta(z-a) \Big] E_1(z)
= (k_\perp^2 - \omega^2) P_1(z) -ik_\perp \frac{\partial}{\partial z} P_3(z),
\quad \text{for} \quad {\bm\lambda}_g={\bf 0}, \quad \lambda_e^{||}=0. 
\label{2order-E1}
\end{equation}
If $E_1(z)$ were to have a $\delta$-function singularity on the plate, 
then the derivatives of $E_1(z)$ would have higher order singularities
and Eq.~(\ref{2order-E1}) can not be consistently balanced.
Thus, the conditions in Eq.~(\ref{sing-con}) are necessary for consistency.

The boundary conditions\footnote{%
These boundary conditions differ from those considered in 
Ref.~\cite{Fosco:2012mtt} which attempts to model a $\delta$-function plate
in the context of relativistic macroscopic electrodynamics.
The boundary conditions we consider here more closely model a 
$\delta$-function plate with nonrelativistic matter inside.
There is no distinction between ${\bm\lambda}_e$ and ${\bm\lambda}_g$
in Ref.~\cite{Fosco:2012mtt}, and although the same Casimir energy
is recovered in the perfect conductor limit, their results in general
do not agree with those obtained in the present article.}
on the TM mode are found, 
by integrating Eqs.~(\ref{TM-first-order}), to be 
\begin{subequations}
\begin{eqnarray}
\lambda_e^{||} E_3(a) &=& 0, \label{TM-bc-D3} \\
D_3(a+\delta) - D_3(a-\delta) &=& -ik_\perp \lambda_e^\perp E_1(a),
\label{TM-bc-D3pm} \\
E_1(a+\delta) - E_1(a-\delta) &=& i\omega \lambda_g^\perp H_2(a),
\end{eqnarray}%
\label{TM-bc}%
\end{subequations}
and the corresponding boundary conditions on the TE mode are found,
by integrating Eqs.~(\ref{TE-first-order}), to be 
\begin{subequations}
\begin{eqnarray}
\lambda_g^{||} H_3(a) &=& 0, \label{TE-bc-H3} \\
B_3(a+\delta) - B_3(a-\delta) &=& -ik_\perp \lambda_g^\perp H_1(a), \\
H_1(a+\delta) - H_1(a-\delta) &=& -i\omega \lambda_e^\perp E_2(a).
\end{eqnarray}%
\label{TE-bc}%
\end{subequations}
Equations~(\ref{TM-bc}) and (\ref{TE-bc}) reduce to the correct boundary
conditions at the interface of two semi-infinite slabs without surface 
charges and currents when the $\delta$-function plate is absent
(obtained by omitting terms involving $\lambda_{e,g}^{\perp,||}$'s).
Because the position of the polarization source in Eqs.~(\ref{ME-cross}) is
at our disposal we can choose it to lie outside the integration region
to derive Eqs.~(\ref{TM-bc}) and (\ref{TE-bc}).
In Eqs.~(\ref{TM-bc}) and (\ref{TE-bc}) some fields have to be 
evaluated on the semitransparent $\delta$-function plate, and
we will see that $E_3(z)$ and $H_3(z)$ in general are discontinuous
at $z=a$. The evaluation of such discontinuous fields at $z=a$ is
mathematically ill-defined. But, when the $\delta$-functions in 
Eq.~(\ref{epmu-def}) are interpreted as a limit of a sequence of 
functions that are symmetric about $z=a$ \cite{Estrada:2007pt}
these fields are readily evaluated 
as the average of their left and right limits at $z=a$ even if these 
limits do not coincide (as long as both limits exist).
This ``averaging prescription'' has been proved in Appendix~\ref{avpres-sec}.
The evaluations in Eqs.~(\ref{TM-bc-D3}) and (\ref{TE-bc-H3})
thus should be understood as averages of the values of the 
discontinuous function on either side of the semitransparent 
$\delta$-function plate.
This averaging prescription  was previously argued on
heuristic grounds \cite{CaveroPelaez:2008tj} and
was successfully employed to calculate the lateral 
Casimir force and Casimir torque between semitransparent $\delta$-function
surfaces with sinusoidal corrugations for planar as well as cylindrical 
geometries in Refs.~\cite{CaveroPelaez:2008tj} and \cite{CaveroPelaez:2008tk}
involving interactions mediated by a scalar field.

Boundary conditions on a semitransparent $\delta$-function plate
in Eqs.~(\ref{TM-bc}) and (\ref{TE-bc}) were also derived
by Barton in Ref.~\cite{Barton:2005eps}. Barton does not explicitly write
Eqs.~(\ref{TM-bc-D3}) and (\ref{TE-bc-H3}) in
Ref.~\cite{Barton:2005eps} but these conditions are implicit in his model.
Equations~(\ref{TM-bc-D3}) and (\ref{TE-bc-H3}) imply that 
$\lambda_e^{||}=0$ unless $E_3(a)=0$, and $\lambda_g^{||}=0$ unless
$H_3(a)=0$. One thus can conclude that a semitransparent
$\delta$-function plate necessarily is anisotropic in its optical properties.
We postpone discussion of the boundary conditions on a 
perfect electrically conducting $\delta$-function plate
until Sec.~\ref{bc-tp-sec}.


\subsection{Fields and Green's functions}
\label{fields-and-gf-sec}

To keep our discussions open to the implications of
Eqs.~(\ref{TM-bc-D3}) and (\ref{TE-bc-H3})
we shall not use isotropic considerations for simplification.
We define the magnetic Green's function $g^H(z,z^\prime)$, and 
the electric Green's function $g^E(z,z^\prime)$, as the inverse of
the differential operators in Eqs.~(\ref{2order-D3B3}), to construct
\begin{subequations}
\begin{eqnarray}
\left[ - \frac{\partial}{\partial z} \frac{1}{\varepsilon^\perp(z)}
\frac{\partial}{\partial z} + \frac{k_\perp^2}{\varepsilon^{||}(z)} 
-\omega^2 \mu^\perp(z) \right] g^H(z,z^\prime) &=& \delta(z-z^\prime), 
\label{greenH} \\
\left[ - \frac{\partial}{\partial z} \frac{1}{\mu^\perp(z)}
\frac{\partial}{\partial z} + \frac{k_\perp^2}{\mu^{||}(z)} 
-\omega^2 \varepsilon^\perp(z) \right] g^E(z,z^\prime) &=& \delta(z-z^\prime), 
\label{greenE}
\end{eqnarray}%
\label{green-funs}%
\end{subequations}
written in terms of anisotropic material properties (with isotropy in the 
plane). 
The components of the fields are obtained in terms of the Green's functions
by inverting Eqs.~(\ref{2order-D3B3}) and the corresponding equations
for other components. These are expressed in the form 
\begin{subequations}
\begin{eqnarray}
{\bf E}(z) 
= \int dz^\prime \,{\bm \gamma} (z,z^\prime) \cdot {\bf P}(z^\prime), \\
{\bf H}(z) 
= \int dz^\prime \,{\bm \phi} (z,z^\prime) \cdot {\bf P}(z^\prime),
\end{eqnarray}%
\label{E=Gp,H=PP}%
\end{subequations}
in terms of the reduced Green's dyadics
\begin{equation}
{\bm\gamma}(z,z^\prime)
= \left[ \begin{array}{ccc}
\frac{1}{\varepsilon^\perp(z)} \frac{\partial}{\partial z}
\frac{1}{\varepsilon^\perp(z^\prime)} \frac{\partial}{\partial z^\prime}
 g^H(z,z^\prime) & 0 &
\frac{1}{\varepsilon^\perp(z)} \frac{\partial}{\partial z}
\frac{ik_\perp}{\varepsilon^{||}(z^\prime)} g^H(z,z^\prime) \\[2mm]
0 & \omega^2 g^E(z,z^\prime) & 0 \\[2mm]
-\frac{ik_\perp}{\varepsilon^{||}(z)} \frac{1}{\varepsilon^\perp(z^\prime)} 
\frac{\partial}{\partial z^\prime} g^H(z,z^\prime) & 0 &
-\frac{ik_\perp}{\varepsilon^{||}(z)}
\frac{ik_\perp}{\varepsilon^{||}(z^\prime)} g^H(z,z^\prime)
\end{array} \right]
-\delta(z-z^\prime)
\left[ \begin{array}{llr}
\frac{1}{\varepsilon^\perp(z)} \hspace{2mm}& 0 \hspace{2mm} & 0 \\
0 & 0 & 0 \\ 0 & 0 & \frac{1}{\varepsilon^{||}(z)} 
\end{array} \right]
\label{Gamma=gE}
\end{equation}
and
\begin{equation}
{\bm\phi}(z,z^\prime)
= i\omega \left[ \begin{array}{ccc}
0 & \frac{1}{\mu^\perp(z)} \frac{\partial}{\partial z} g^E(z,z^\prime)
& 0 \\[2mm]
\frac{1}{\varepsilon^\perp(z^\prime)}
\frac{\partial}{\partial z^\prime} g^H(z,z^\prime) & 0 &
\frac{ik_\perp}{\varepsilon^{||}(z^\prime)} g^H(z,z^\prime) \\[2mm]
0 & -\frac{ik_\perp}{\mu^{||}(z)} g^E(z,z^\prime) & 0
\end{array} \right].
\label{Phi=gH}
\end{equation}
These are straightforward generalizations of the Green's dyadics
in Ref.~\cite{Schwinger:1977pa} to the anisotropic case. 
Since we have already Fourier transformed in the $xy$-plane the
Green's dyadics in Eqs.~(\ref{Gamma=gE}) and (\ref{Phi=gH}) are
defined in the Fourier space. For example, the electric Green's dyadic
is defined as
\begin{equation}
{\bm\Gamma}({\bf r},{\bf r}^\prime;\omega)
= \int \frac{d^2k_\perp}{(2\pi)^2} 
\,e^{i{\bf k}_\perp \cdot ({\bf r}-{\bf r}^\prime)_\perp} 
{\bm\gamma}(z,z^\prime;{\bf k}_\perp,\omega).
\end{equation}
The solutions to Eqs.~(\ref{green-funs}) 
for the magnetic and electric Green's functions
completely determine the fields by Eqs.~(\ref{E=Gp,H=PP}).
In other words, the boundary conditions for the fields on a
semitransparent $\delta$-function plate thus impose conditions on 
the magnetic and electric Green's functions at the position of the plate.


\section{Magnetic and Electric Green's functions}
\label{HEGreen}

\begin{center} \begin{figure}
\includegraphics[width=6.5cm]{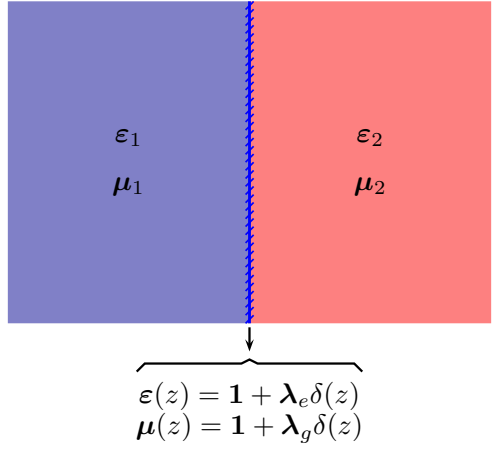}
\caption{A semitransparent $\delta$-function plate sandwiched between two
semi-infinite slabs with the material properties of Eq.~(\ref{epmu-ds}).}
\label{slabs-delta-fig}
\end{figure} \end{center}

In this section we find solutions to the magnetic and electric 
Green's functions introduced in Eqs.~(\ref{green-funs}) and satisfying
boundary conditions dictated by Eqs.~(\ref{TM-bc}) and (\ref{TE-bc}).
Since we do not restrict our discussion to $\lambda^{||}_{e,g}=0$, our
solutions in this section could be used to analyze the alternative choices
$E_3(a)=0$ or $H_3(a)=0$ in Eqs.~(\ref{TM-bc-D3}) and (\ref{TE-bc-H3}).
The treatment in this section considers a $\delta$-function plate
sandwiched between two uniaxial materials, described by 
\begin{subequations}
\begin{eqnarray}
\bm{\varepsilon} (z) &=& \varepsilon^\perp (z)\, {\bf 1}_\perp 
+ \varepsilon^{||}(z) \, \hat{\bf z} \,\hat{\bf z}, \\
\bm{\mu} (z) &=& \mu^\perp (z)\, {\bf 1}_\perp 
+ \mu^{||} (z) \,\hat{\bf z} \,\hat{\bf z},
\end{eqnarray}%
\label{epmu-ds}
\end{subequations}
where
\begin{subequations}
\begin{eqnarray}
\varepsilon^{\perp,||} (z) &=& 1 + (\varepsilon^{\perp,||}_1-1) \theta(a-z) 
+ (\varepsilon^{\perp,||}_2-1) \theta(z-a) + \lambda^{\perp,||}_e \delta(z-a),
\\
\mu^{\perp,||} (z) &=& 1 + (\mu^{\perp,||}_1-1) \theta(a-z) 
+ (\mu^{\perp,||}_2-1) \theta(z-a) + \lambda^{\perp,||}_g \delta(z-a).
\end{eqnarray}%
\label{eps-def}%
\end{subequations}
We have used Heaviside step functions, or $\theta$-functions,
\begin{equation}
\theta(z) = \begin{cases} 1 & \text{if} \quad z>0, \\
0 & \text{if} \quad z<0, \end{cases}
\label{th-f-def}
\end{equation}
and $\delta$-functions in Eqs.~(\ref{eps-def})
to describe discontinuities and singularities in the material properties.
See Fig.~\ref{slabs-delta-fig}. Setting 
$\varepsilon^\perp_i=\varepsilon^{||}_i=1$ and $\mu^\perp_i=\mu^{||}_i=1$ 
corresponds to a semitransparent $\delta$-function plate in vacuum.

The boundary conditions on the fields of 
Eqs.~(\ref{TM-bc}) and (\ref{TE-bc}) are transcribed onto the Green's
dyadics through Eqs.~(\ref{E=Gp,H=PP}).
The boundary conditions on the TM mode in Eqs.~(\ref{TM-bc})
give the conditions
\begin{subequations}
\begin{eqnarray}
\varepsilon^{||}_2 \,\gamma_{3i}(a+\delta,z^\prime)
-\varepsilon^{||}_1 \,\gamma_{3i}(a-\delta,z^\prime)
&=& - ik_\perp \lambda^\perp_e \frac{1}{2}
\big[ \gamma_{1i}(a+\delta,z^\prime) + \gamma_{1i}(a-\delta,z^\prime) \big],
\label{Gamma-bc-a} \\
\gamma_{1i}(a+\delta,z^\prime) - \gamma_{1i}(a-\delta,z^\prime)
&=& i \omega \lambda^\perp_g \frac{1}{2}
\big[ \phi_{2i}(a+\delta,z^\prime) + \phi_{2i}(a-\delta,z^\prime) \big],
\end{eqnarray}%
\label{Gamma-bc}%
\end{subequations}
and the corresponding boundary conditions on the TE mode in Eqs.~(\ref{TE-bc})
gives
\begin{subequations}
\begin{eqnarray}
\mu^{||}_2 \,\phi_{3i}(a+\delta,z^\prime)
-\mu^{||}_1 \,\phi_{3i}(a-\delta,z^\prime)
&=& - ik_\perp \lambda^\perp_g \frac{1}{2}
\big[ \phi_{1i}(a+\delta,z^\prime) + \phi_{1i}(a-\delta,z^\prime) \big], 
\label{Phi-bc-a} \\
\phi_{1i}(a+\delta,z^\prime) - \phi_{1i}(a-\delta,z^\prime)
&=& -i \omega \lambda^\perp_e \frac{1}{2}
\big[ \gamma_{2i}(a+\delta,z^\prime) + \gamma_{2i}(a-\delta,z^\prime) \big].
\end{eqnarray}%
\label{Phi-bc}%
\end{subequations}
Using Eqs.~(\ref{Gamma=gE}) and (\ref{Phi=gH}),
the boundary conditions in Eqs.~(\ref{Gamma-bc}) and (\ref{Phi-bc})
on the Green's dyadics in turn dictate
the boundary conditions on the electric and magnetic Green's functions.
These are more efficiently found by setting $i=3$ in 
Eqs.~(\ref{Gamma-bc}) to obtain the boundary conditions on the 
magnetic Green's function,
\begin{subequations}
\begin{eqnarray}
g^H(z,z^\prime) \Big|^{z=a+\delta}_{z=a-\delta}
&=& \frac{\lambda^\perp_e}{2}
\left[ \left\{ \frac{1}{\varepsilon^\perp(z)} \frac{\partial}{\partial z}
g^H(z,z^\prime) \right\}_{z=a+\delta} 
+ \left\{ \frac{1}{\varepsilon^\perp(z)} \frac{\partial}{\partial z}
g^H(z,z^\prime) \right\}_{z=a-\delta} \right], \\
\left\{ \frac{1}{\varepsilon^\perp(z)} \frac{\partial}{\partial z}
g^H(z,z^\prime) \right\}\bigg|^{z=a+\delta}_{z=a-\delta}
&=& \zeta^2 \frac{\lambda^\perp_g}{2}
\big[ g^H(a+\delta,z^\prime) + g^H(a-\delta,z^\prime) \big].
\end{eqnarray}%
\label{gH-bc}%
\end{subequations} 
Similarly, setting $i=2$ in Eqs.~(\ref{Phi-bc}) yields the boundary 
conditions on the electric Green's function,
\begin{subequations}
\begin{eqnarray}
g^E(z,z^\prime) \Big|^{z=a+\delta}_{z=a-\delta}
&=& \frac{\lambda^\perp_g}{2}
\left[ \left\{ \frac{1}{\mu^\perp(z)} \frac{\partial}{\partial z}
g^E(z,z^\prime) \right\}_{z=a+\delta} 
+ \left\{ \frac{1}{\mu^\perp(z)} \frac{\partial}{\partial z}
g^E(z,z^\prime) \right\}_{z=a-\delta} \right], \\
\left\{ \frac{1}{\mu^\perp(z)} \frac{\partial}{\partial z}
g^E(z,z^\prime) \right\}\bigg|^{z=a+\delta}_{z=a-\delta}
&=& \zeta^2 \frac{\lambda^\perp_e}{2}
\big[ g^E(a+\delta,z^\prime) + g^E(a-\delta,z^\prime) \big].
\end{eqnarray}%
\label{gE-bc}%
\end{subequations}
Note that we have switched to imaginary frequencies by performing
the Euclidean rotation, $\omega \to i\zeta$.

The solution for the magnetic Green's function satisfying the 
boundary conditions in Eqs.~(\ref{gH-bc}) is
\begin{equation}
g^H(z,z^\prime) = 
\begin{cases}
\frac{1}{2\bar{\kappa}^H_1} \Big[ e^{-\kappa^H_1 |z-z^\prime|}
+ r^H_{12} \, e^{-\kappa^H_1 |z-a|} e^{-\kappa^H_1 |z^\prime-a|} \Big], 
& \text{if} \quad z,z^\prime < a, \\[2mm]
\frac{1}{2\bar{\kappa}^H_2} \Big[ e^{-\kappa^H_2 |z-z^\prime|}
+ r^H_{21} \, e^{-\kappa^H_2 |z-a|} e^{-\kappa^H_2 |z^\prime-a|} \Big], 
& \text{if} \quad a < z,z^\prime, \\[2mm]
\frac{1}{2\bar{\kappa}^H_2} \, t^H_{21} \, 
e^{-\kappa^H_1 |z-a|} e^{-\kappa^H_2 |z^\prime-a|}, 
& \text{if} \quad z<a < z^\prime, \\[2mm]
\frac{1}{2\bar{\kappa}^H_1} \,t^H_{12} \,
e^{-\kappa^H_2 |z-a|} e^{-\kappa^H_1 |z^\prime-a|}, 
& \text{if} \quad z^\prime <a <z, 
\end{cases}
\label{gH-sol}
\end{equation}
where the reflection coefficients are
\begin{equation}
r^H_{ij} =
\frac{\bar{\kappa}^H_i 
\Big( 1 + \frac{\lambda^\perp_e \bar{\kappa}^H_j}{2} \Big)
\Big( 1 - \frac{\lambda^\perp_g \zeta^2}{2\bar{\kappa}^H_i} \Big)
-\bar{\kappa}^H_j 
\Big( 1 - \frac{\lambda^\perp_e \bar{\kappa}^H_i}{2} \Big)
\Big( 1 + \frac{\lambda^\perp_g \zeta^2}{2\bar{\kappa}^H_j} \Big) }
{ \bar{\kappa}^H_i 
\Big( 1 + \frac{\lambda^\perp_e \bar{\kappa}^H_j}{2} \Big)
\Big( 1 + \frac{\lambda^\perp_g \zeta^2}{2\bar{\kappa}^H_i} \Big)
+\bar{\kappa}^H_j 
\Big( 1 + \frac{\lambda^\perp_e \bar{\kappa}^H_i}{2} \Big)
\Big( 1 + \frac{\lambda^\perp_g \zeta^2}{2\bar{\kappa}^H_j} \Big) },
\label{rHij}
\end{equation}
and the transmission coefficients are
\begin{equation}
t^H_{ij} =
\frac{\bar{\kappa}^H_i 
\Big( 1 + \frac{\lambda^\perp_e \bar{\kappa}^H_i}{2} \Big)
\Big( 1 - \frac{\lambda^\perp_g \zeta^2}{2\bar{\kappa}^H_i} \Big)
+\bar{\kappa}^H_i 
\Big( 1 - \frac{\lambda^\perp_e \bar{\kappa}^H_i}{2} \Big)
\Big( 1 + \frac{\lambda^\perp_g \zeta^2}{2\bar{\kappa}^H_i} \Big) }
{ \bar{\kappa}^H_i 
\Big( 1 + \frac{\lambda^\perp_e \bar{\kappa}^H_j}{2} \Big)
\Big( 1 + \frac{\lambda^\perp_g \zeta^2}{2\bar{\kappa}^H_i} \Big)
+\bar{\kappa}^H_j 
\Big( 1 + \frac{\lambda^\perp_e \bar{\kappa}^H_i}{2} \Big)
\Big( 1 + \frac{\lambda^\perp_g \zeta^2}{2\bar{\kappa}^H_j} \Big) },
\label{tHij}
\end{equation}
with
\begin{equation}
\kappa_i^H
= \sqrt{k_\perp^2 \frac{\varepsilon^\perp_i}{\varepsilon^{||}_i}
+ \zeta^2 \varepsilon^\perp_i \mu^\perp_i}
\qquad \text{and} \qquad
\bar{\kappa}^H_i = \frac{\kappa^H_i}{\varepsilon^\perp_i} 
=\sqrt{ \frac{k_\perp^2}{\varepsilon^\perp_i \varepsilon^{||}_i}
+ \zeta^2 \frac{\mu^\perp_i}{\varepsilon^\perp_i} }.
\label{kaiH}
\end{equation}
The electric Green's function is obtained from the magnetic Green's 
function by replacing ${\bm \varepsilon} \leftrightarrow {\bm \mu}$
and $H\to E$, with the corresponding definitions
\begin{equation}
\kappa_i^E
= \sqrt{k_\perp^2 \frac{\mu^\perp_i}{\mu^{||}_i}
+ \zeta^2 \mu^\perp_i \varepsilon^\perp_i}
\qquad \text{and} \qquad
\bar{\kappa}^E_i = \frac{\kappa^E_i}{\mu^\perp_i}
=\sqrt{ \frac{k_\perp^2}{\mu^\perp_i \mu^{||}_i}
+ \zeta^2 \frac{\varepsilon^\perp_i}{\mu^\perp_i} }.
\label{kaiE}
\end{equation}
Setting $\lambda^\perp_e=\lambda^\perp_g=0$, Eqs.~(\ref{rHij})
and (\ref{tHij}) immediately lead to the standard reflection and
transmission coefficients at the interface of two semi-infinite slabs,
and serve as a check for the reflection and transmission coefficients
in Eqs.~(\ref{rHij}) and (\ref{tHij}). 

It should be emphasized that even though we explicitly considered 
materials with $\lambda^{||}_e$ and  $\lambda^{||}_g$  
in Eqs.~(\ref{eps-def}), the solutions to the 
Green's functions given by Eqs.~(\ref{gH-sol}) through (\ref{kaiE})
are independent of $\lambda^{||}_e$ and  $\lambda^{||}_g$.
The Green's functions of Eqs.~(\ref{gH-sol}) determine the fields
unambiguously everywhere (except on the $\delta$-function plate)
and the implication is that there are no observable consequences
of $\lambda^{||}_e$ and $\lambda^{||}_g$.

\subsection{Green's functions for a semitransparent $\delta$-function plate}
 
A semitransparent $\delta$-function plate in vacuum corresponds to setting
$\varepsilon^\perp_i=\varepsilon^{||}_i=1$ and $\mu^\perp_i=\mu^{||}_i=1$.
This simplifies the expressions for the reflection and transmission
coefficients in Eqs.~(\ref{rHij}) and (\ref{tHij}) significantly because
$\kappa_i^H$ and $\kappa_i^E$ no longer are distinct. In terms of 
\begin{equation}
\kappa^2 = k_\perp^2 + \zeta^2,
\end{equation}
the magnetic Green's function for a semitransparent $\delta$-function plate
in vacuum can be expressed in the compact form 
\begin{equation}
g^H(z,z^\prime) = \frac{1}{2\kappa} e^{-\kappa |z-z^\prime|}
+ \big[ r^H_g + \eta(z-a) \eta(z^\prime -a) \, r^H_e \big]
\frac{1}{2\kappa} e^{-\kappa |z-a|} e^{-\kappa |z^\prime -a|},
\label{gHtp}
\end{equation}
where
\begin{equation}
\eta(z) = \begin{cases} 1, & \text{if} \quad z>0, \\
-1, & \text{if} \quad z<0. \end{cases}
\label{eta-fun}
\end{equation}
Equation~(\ref{gHtp}) is written in terms of contributions
to the reflection coefficients of the transverse magnetic
mode for the two special cases of vanishing electric permittivity
and vanishing magnetic permeability.
These reflection coefficients, $r^H_e$ and $r^H_g$,
and the corresponding transmission coefficients, $t^H_e$ and $t^H_g$,
are related by,
\begin{equation}
r^H_e = \frac{\lambda^\perp_e}{\lambda^\perp_e + \frac{2}{\kappa}},
\quad t^H_e = 1 - r^H_e,
\qquad \text{and} \qquad
r^H_g = - \frac{\lambda^\perp_g}{\lambda^\perp_g + \frac{2\kappa}{\zeta^2}},
\quad t^H_g = 1 + r^H_g.
\label{rHeg}
\end{equation}
The total reflection and transmission coefficients for the magnetic mode
(easily read off from Eq.~(\ref{gHtp}) with reference to Eqs.~(\ref{gH-sol}))
are 
\begin{equation}
r^H=r^H_g+r^H_e, \qquad t^H=1+r^H_g - r^H_e.
\label{rHtH-def}
\end{equation}
In terms of these reflection coefficients the boundary conditions
for the magnetic Green's function in Eqs.~(\ref{gH-bc}) read
\begin{subequations}
\begin{eqnarray}
\big[ 1+r^H-t^H \big] 
&=& \frac{\lambda_e^\perp\kappa}{2} \big[ 1-r^H+t^H \big], \\
\big[ 1-r^H-t^H \big] 
&=& \frac{\lambda_g^\perp\zeta^2}{2\kappa} \big[ 1+r^H+t^H \big].
\end{eqnarray}%
\label{rtH-bc}%
\end{subequations}
Combining Eqs.~(\ref{rtH-bc}) is expected to yield the statement 
of conservation of energy for the TM-mode \cite{Schwinger:1998cla},
\begin{equation}
1 - (r^H)^2 - (t^H)^2 
= \frac{1}{2} \frac{\lambda_e^\perp\kappa}{2} \big[ 1-r^H+t^H \big]^2
+ \frac{1}{2} \frac{\lambda_g^\perp\zeta^2}{2\kappa} 
\big[ 1+r^H+t^H \big]^2.
\label{econH}
\end{equation}
This calls for an investigation of conservation of energy in the 
presence of a semitransparent $\delta$-function plate.
From Maxwell's equations in Eqs.~(\ref{ME-cross})
the local statement of conservation of energy is
\begin{equation}
{\bm \nabla} \cdot ({\bf E} \times {\bf H}) 
+ {\bf E} \cdot \frac{\partial}{\partial t} {\bf D}
+ {\bf H} \cdot \frac{\partial}{\partial t} {\bf B} =0.
\label{eflux}
\end{equation}
It equates the energy flux into a volume with the rate of change
of the energy density. When integrated across the
$\delta$-function plate in the $\hat{\bf z}$-direction this yields
Eq.~(\ref{econH}) and thus verifies that our description by 
a $\delta$-function plate conserves energy.
It is to be noted that the terms on the left hand side in 
Eq.~(\ref{econH}) come from the energy flux term in Eq.~(\ref{eflux}),
and the terms on the right hand side are contributed by the energy density.

From the general solution in Eq~(\ref{gHtp})
one can read off the contributions for 
the special case when the $\delta$-function plate is purely
electric ($r_g^H=0$),
or when the $\delta$-function plate is purely magnetic ($r_e^H =0$).
A perfect electrical conductor is a material with $r^H_e=1$ and $r^H_g=0$.
This is found by taking
the limit $\lambda^\perp_e \to\infty$ and $\lambda^\perp_g =0$
in Eq.~(\ref{rHeg}). Similarly, a perfect magnetic conductor, described by 
$r^H_e=0$ and $r^H_g=-1$, is obtained when 
$\lambda^\perp_e =0$ and $\lambda^\perp_g \to \infty$.
A perfect electric and magnetic conductor interestingly is described by 
$r^H_e=1$ and $r^H_g=-1$. This corresponds to $r^H=0$ and $t^H=-1$,
and implies the total transmission of an incident wave with 
a phase shift of $\pi$ when crossing the $\delta$-function plate.
Insertion of such a plate in a circular laser would extinguish it
if the circumference equals an integer multiple of the wavelength
of the laser.

Similar expressions hold for the electric Green's function
obtained by swapping $\lambda_e^\perp \leftrightarrow \lambda_g^\perp$
and replacing superscripts $H \rightarrow E$.

\subsection{Boundary conditions on a perfect electrically 
conducting $\delta$-function plate}
\label{bc-tp-sec}

Before we consider a perfect electrically conducting $\delta$-function plate,
let us first consider the boundary conditions at the interface of
two semi-infinite dielectric slabs with 
${\bm \varepsilon}_1 \ll {\bm \varepsilon}_2$, with no $\delta$-function plate
sandwiched in between, ${\bm\lambda}_e={\bm\lambda}_g=0$.
Boundary conditions on the TM-mode in Eqs.~(\ref{TM-bc}) for this case read
$\varepsilon^{||}_2 E_3(a+\delta) = \varepsilon^{||}_1 E_3(a-\delta)$,
and $E_1(a+\delta) = E_1(a-\delta)$, with no information content 
in Eqs.~(\ref{TM-bc-D3}).
Using the solutions for the Green's functions in Eq.~(\ref{gH-sol})
we can evaluate the electric fields using Eqs.~(\ref{E=Gp,H=PP}). For 
convenience let us choose the polarization source to be a
polarized point source
${\bf P} \propto (1,0,0) \, \delta^{(3)}({\bf x}-{\bf x}^\prime)$
outside the interface.
With the intention of taking the limit of infinite dielectric response
in the slab on the right, ${\bm \varepsilon}_2\to\infty$, we choose the
source point just to the left,
$z^\prime=a-\delta$, because it is meaningless to have an electromagnetic
source inside a perfect conductor. In this manner we can show that:
$ E_1(a+\delta) = E_1(a-\delta) = - \bar{\kappa}_1^H (1-r_{12}^H)/2$, and
$\varepsilon^{||}_2 E_3(a+\delta) = \varepsilon^{||}_1 E_3(a-\delta)
= -ik_\perp (1+r_{12}^H)/2$.
For the case when the second medium is perfectly conducting in the
$xy$-plane, we have $\varepsilon_2^\perp \to\infty$ and $r_{12}^H \to 1$,
which implies:
$E_1(a+\delta) = E_1(a-\delta)=0$, and
$\varepsilon_2^{||} E_3(a+\delta) = \varepsilon_1^{||} E_3(a-\delta)
= -ik_\perp$.
Thus perfect conduction in $xy$-plane alone sets the tangential 
component of electric field to zero, but does not require the 
normal component of electric field to be zero.
If we further require infinite conductance in the $z$-direction,
we have $\varepsilon_2^{||} \to\infty$, which implies:
$E_3(a-\delta) = -ik_\perp/\varepsilon_1^{||}$ and
$E_3(a+\delta)=0$.
These are the boundary conditions satisfied by the electric field
on a perfect isotropic conductor.
Notice the interesting fact that with the normal component of the electric field
being zero inside the conductor and with infinite dielectric 
response in the $z$-direction inside the conductor,
we still have a well defined finite value for the normal component
of the electric displacement field,
$\lim_{\varepsilon_2^\perp \to\infty,\,
\varepsilon_2^{||}\to\infty} D_3(a+\delta) =-ik_\perp$.
This is necessary to satisfy the continuity condition for the 
normal component of the macroscopic field at the interface.

Let us next consider a semitransparent electric $\delta$-function plate
in vacuum by setting ${\bm \varepsilon}_1 = {\bm \varepsilon}_2 =1$ and 
${\bm \lambda}_g=0$. In principle now there is no need to restrict
the source point to $z^\prime=a-\delta$. For this case we can evaluate
\begin{equation}
E_1(a) = E_1(a+\delta) = E_1(a-\delta) = - \frac{\kappa}{2} (1-r_e^H)
\xrightarrow{\lambda_e^\perp \to\infty} 0,
\label{E1-pecdp}
\end{equation}
where we used the limiting condition, $r_e^H\to 1$, using Eq.~(\ref{rHeg}).
Similarly, we can evaluate, for $z^\prime=a-\delta$,
\begin{subequations}
\begin{eqnarray}
E_3(a+\delta) &=& 
-\frac{ik_\perp}{2} (1-r_e^H) \xrightarrow{\lambda_e^\perp \to\infty} 0, \\ 
E_3(a-\delta) &=& 
-\frac{ik_\perp}{2}(1+r_e^H) \xrightarrow{\lambda_e^\perp \to\infty} -ik_\perp. 
\end{eqnarray}%
\label{E3-per-con}%
\end{subequations}
When the source is placed at an arbitrary distance $z^\prime$ all
the above expressions contain an exponentially decaying factor
of $\exp(-\kappa |z^\prime-a|)$.
We further note that the factor contributing on the right 
hand side of Eq.~(\ref{TM-bc-D3pm}) is
\begin{equation}
\lambda_e^\perp E_1(a) = - r_e^H \xrightarrow{\lambda_e^\perp \to\infty} -1. 
\label{laE1=-1}
\end{equation}
Thus, there is
no distinction between a semi-infinite perfect isotropic electric conductor
and a perfect electrically conducting $\delta$-function plate.
To clarify how these two cases conspire to yield the 
same physics in the perfect conducting limit
we rewrite Eq.~(\ref{TM-bc-D3pm}) in the form
\begin{equation}
D_3(a-\delta) = D_3(a+\delta) + ik_\perp \lambda_e^\perp E_1(a)
= \begin{cases} -ik_\perp + 0, & \text{for} \quad {\bm\lambda}_e={\bf 0}, 
\quad \hspace{2.8mm} {\bm \varepsilon}_2=\varepsilon_2{\bf 1} \to \infty, \\
0 - ik_\perp, & \text{for} \quad \lambda_e^\perp \to \infty, \quad
{\bm \varepsilon}_2={\bf 1},
\end{cases}
\end{equation}
where we explicitly note that while for a 
semi-infinite perfect isotropic electric conductor the 
contribution on the right is from $D_3(a+\delta)$,
for a perfect electrically conducting $\delta$-function plate the
contribution is from the term contributing on the $\delta$-function plate.

Similar analysis for the TE-mode leads to the corresponding boundary 
conditions on the magnetic fields.


\section{Physical realization of a semitransparent $\delta$-function plate}
\label{physi-dp}

From this point on, to simplify the analysis, we consider a purely
electrical material by considering the special case $\lambda_g^\perp=0$.
Further, we shall emphasize the frequency dependence by writing
$\lambda_e^\perp(i\zeta)$.

In Refs.~\cite{Barton:2005eps,Barton:2005fst} Barton considers an
infinitesimally thin plasma sheet carrying a continuous fluid
with surface charge and current densities modelled in terms of the
fluid displacement which is assumed to be purely tangential.
Barton's hydrodynamical model is identical to our consideration here
if we require the frequency response to be
\begin{equation}
\lambda_e^\perp(i\zeta) = \frac{\zeta_p}{\zeta^2},
\label{lame=pl}
\end{equation}
where the parameter $\zeta_p$ corresponds to the characteristic
wavenumber in Barton's model~\cite{Barton:2005eps}.


\subsection{Thin-plate limit}
\label{tpl-sec}

We begin by inquiring, in what approximation will a dielectric
slab of thickness $d$ simulate a (purely electric) 
semitransparent $\delta$-function plate? To this end we write the
$\delta$-function in Eq.~(\ref{epmu-def-e}) in terms of $\theta$-functions, 
\begin{equation}
\varepsilon^{\perp,||}(i\zeta) -1 = \lambda_e^{\perp,||}(i\zeta)
\lim_{d\to 0}\frac{[\theta(z+d)-\theta(z)]}{d},
\label{e-1=lz2d}
\end{equation}
which in the limit $d\to 0$ exactly yields Eq.~(\ref{epmu-def-e}).
Equation~(\ref{e-1=lz2d}) describes a dielectric
slab of thickness $d$ if we read the factor $\lambda_e^{\perp,||}(i\zeta)/d$ 
to represent the slab's susceptibility.
 
One can find the reflection coefficients for the TM--and TE-modes
for the dielectric slab described by Eq.~(\ref{e-1=lz2d})
after solving for the corresponding Green's functions in Eqs.~(\ref{green-funs}),
see for example Ref.~\cite{Prachi:2011ec}.
We show that when the following conditions for the thin-plate limit are met,
\begin{equation}
\zeta^2 \ll \frac{\zeta_p}{d} \ll \frac{1}{d^2}, 
\qquad \text{and} \qquad
k_\perp^2 \ll \frac{\zeta_p}{d} \ll \frac{1}{d^2}, 
\label{dtp-con}
\end{equation}
the reflection coefficients for the TM--and TE-modes for a dielectric 
slab of thickness $d$ have the following limiting behavior 
\cite{Prachi:2011ec},
\begin{subequations}
\begin{eqnarray}
r_\text{thick}^H 
&=& -\left(\frac{\bar{\kappa}_i^H -\kappa}{\bar{\kappa}_i^H+\kappa} \right)
\frac{(1- e^{-2\kappa_i^Hd})} {\left[ 1- 
\left(\frac{\bar{\kappa}_i^H -\kappa}{\bar{\kappa}_i^H+\kappa} \right)^2
e^{-2\kappa_i^Hd}\right]} 
\xrightarrow[k_\perp d \ll \sqrt{\zeta_p d} \ll 1]
{\zeta d \ll \sqrt{\zeta_p d} \ll 1} 
r_e^H = \frac{\lambda_e^\perp(i\zeta)}
{\lambda_e^\perp(i\zeta)+\frac{2}{\kappa}}, \label{rthickH-lim} \\
r_\text{thick}^E 
&=& -\left(\frac{\kappa_i^E -\kappa}{\kappa_i^E+\kappa} \right)
\frac{(1- e^{-2\kappa_i^Ed})} {\left[ 1- 
\left(\frac{\kappa_i^E -\kappa}{\kappa_i^E+\kappa} \right)^2
e^{-2\kappa_i^Ed}\right]} 
\xrightarrow[k_\perp d \ll \sqrt{\zeta_p d} \ll 1]
{\zeta d \ll \sqrt{\zeta_p d} \ll 1}
r_e^E = -\frac{\lambda_e^\perp(i\zeta)}
{\lambda_e^\perp(i\zeta)+\frac{2\kappa}{\zeta^2}}, \label{rthickE-lim}
\end{eqnarray}%
\label{thick-rs}%
\end{subequations}
where $r_e^H$ and $r_e^E$ are the corresponding reflection coefficients
in Eq.(\ref{rHeg}) for a purely electric $\delta$-function plate.
The variables $\kappa_i^H$ and $\kappa_i^E$
in Eqs.~(\ref{thick-rs}) are given by setting 
$\mu^\perp=\mu^{||}=1$ in Eqs.~(\ref{kaiH}) and (\ref{kaiE}).
The above limiting behavior is derived in the following manner.
Using Eq.~(\ref{e-1=lz2d})
we have the dielectric permittivity of the slab given by
$(\varepsilon^\perp -1)=1/x$ in terms of $x= d/\lambda^\perp_e$.
We then consider the approximations
\begin{subequations}
\begin{eqnarray}
-\frac{\bar{\kappa}_i^H -\kappa}{\bar{\kappa}_i^H+\kappa}
&=& 1 - 2 c\sqrt{x} + {\cal O}(x^{3/2}), \\
e^{-2\kappa_i^Hd} &=& 1 - 2\kappa c\sqrt{x} \,\frac{\zeta_p}{\zeta^2}
+ {\cal O} (x^{3/2},\zeta\sqrt{\lambda^\perp_e d}),
\end{eqnarray}%
\label{al-ex-appr}%
\end{subequations}
which then implies the limiting behavior in Eq.~(\ref{rthickH-lim}).
Similarly one derives Eq.~(\ref{rthickE-lim}).
The factor $c=\bar{\kappa}_i^H\sqrt{\varepsilon^\perp}/\kappa$ 
in Eqs.~(\ref{al-ex-appr}) cancels out in the limiting behavior.
The approximations $x\ll 1$ and $\zeta^2 \lambda^\perp_e d\ll 1$ 
necessary for the 
approximations in Eqs.~(\ref{al-ex-appr}) to be valid are the 
approximations for the thin-plate limit which was presented in 
Eq.~(\ref{dtp-con}) in a compact form.
Thus, we have shown that the reflection coefficients of a 
dielectric slab reduce to those of a $\delta$-function plate
in the thin-plate limit of Eq.~(\ref{dtp-con}).
The limiting behavior in Eqs.~(\ref{thick-rs}) 
is a straightforward generalization of the analysis
for the isotropic case carried out in Ref.~\cite{Prachi:2011ec},
and should be contrasted
with setting $d=0$ in $r_\text{thick}^H$ and $r_\text{thick}^E$,
which would imply zero reflectance and suggest non-existence of the plate
in this limit. Note that the reflection coefficients for the thick slab
on the left hand side of Eqs.~(\ref{thick-rs})
contains $\varepsilon^{||}$ inside 
the variables $\kappa_i^H$ and $\kappa_i^E$, but the limiting 
behavior on the right hand side of Eqs.~(\ref{thick-rs}) is 
completely unaware of any dependence on $\varepsilon^{||}$.
Thus, in the thin-plate limit of Eq.~(\ref{dtp-con}) the optical
properties only depend on the transverse properties of the plate.


\subsection{Plasma sheet}

Combining Eq.~(\ref{lame=pl}), with the dielectric model in
Eq.~(\ref{e-1=lz2d}) for the case of a plasma model,
we can identify the plasma frequency,
$\omega_p^2=n_fe^2/m=\zeta_p/d$, for the thin plasma sheet,
where $n_f$ is the number density of charge carriers with 
charge $e$ and mass $m$. The implication seems to be that the 
number density is inversely proportional to the thickness of the material
for a very thin plasma sheet. To analyze this we model the charge carriers
as a non-relativistic Fermi gas,
confined in a slab of thickness $d$ that is infinite in extent along the 
$xy$-directions, with energy states
\begin{equation}
E_n({\bf k}_\perp) = \frac{1}{2m} 
\left[ {\bf k}_\perp^2 + n^2\frac{\pi^2}{d^2} \right], \qquad n=0,1,2,\ldots,
\label{En=k2n2}
\end{equation}
where $n$ represents the discretization due to confinement
in the $z$-direction. Notice that $n=0$ state is not excluded because
one imposes the Neumann boundary condition, which is necessary to have no
probability flux across the walls of the slab. The total number of
charge carriers, $n_\text{tot}$, in the slab is  obtained by 
summing over all the occupied energy states.
Thus, in terms of the Fermi energy, $E_F=k_F^2/2m$, one evaluates
the number density, 
\begin{equation}
n_f=\frac{n_\text{tot}}{A\,d} = 2\,\frac{1}{d} \sum_{n=0}^\infty 
\int_{-\infty}^\infty \frac{dk_x}{2\pi} \int_{-\infty}^\infty \frac{dk_y}{2\pi}
\,\theta (E_F-E_n(k_x,k_y)) = \frac{k_F^3}{3\pi^2} \nu(x),
\label{neb=ne-ex}
\end{equation}
where $A$ is the area of the plate, and
the factor 2 in the second equality is introduced to
accommodate two Fermi charges in each energy state.
The thickness dependence of the number density is captured in the function
\begin{equation}
\nu(x) = \frac{3}{2} \left( x - \frac{1}{3}x^3 \right)
+ \frac{3}{2N} \left( 1 - \frac{1}{2} x^2 \right)
- \frac{1}{4N^2} x,
\label{nu-x}
\end{equation}
\begin{center} \begin{figure}
\includegraphics[width=7cm]{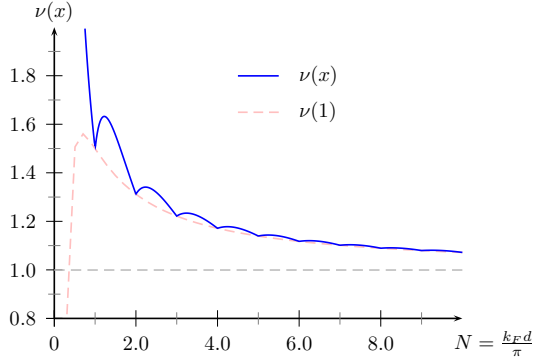}
\caption{Plot of $\nu(x)$ in Eq.~(\ref{nu-x}) versus $N=k_Fd/\pi$.
The value of $\nu(1)$ and of $N\to\infty$ limit are also shown.}
\label{nux-v-x-fig}
\end{figure} \end{center}%
where $x=[N]/N$ is the fractional floor function, $[N]$ referring
to the integer part of $N=k_Fd/\pi$.
For a sufficiently thick slab the difference between $[N]$ and $N$ is
negligible, thus $x\to 1$ and $\nu(x)\to 1$.
A thin plate corresponds to small values of $N$, and in particular 
for $N<1$ we have $\nu(x) \to 3/(2N)$.
The function $\nu(x)$ has been plotted in Fig.~\ref{nux-v-x-fig}.
The limiting cases of the function $\nu(x)$ determine the 
number density in Eq.~(\ref{neb=ne-ex}) for these situations
\cite{Fetter:1973slg,Ashcroft:1976},
\begin{eqnarray}
n_f=\frac{n_\text{tot}}{A\,d} &\to& \begin{cases}
\frac{k_F^2}{2\pi d} & \text{if} \quad N<1 \hspace{5mm}
(\text{2-D sheet}), \\[1mm]
\frac{k_F^3}{3\pi^2} & \text{if} \quad N\to\infty ~~(\text{3-D bulk}).
\end{cases}
\label{ne-lim}
\end{eqnarray}
It is suggestive to explore the physical nature of the condition $N<1$,
which states $d^2 n_\text{tot}/A < \pi/2$.
Introducing the packing fraction of a material defined as
$\nu_0= N_\text{atoms} A_\text{atom}/A$, in terms of the total 
number of atoms $N_\text{atoms}$, and area of an atom 
$A_\text{atom}=\pi a_0^2$ with $a_0$ being the radius of the atom, 
we recognize that the condition $N<1$ corresponds to the window 
$1\leq d/2a_0 < \pi/\sqrt{8\nu_0 n_0} \sim 1$,
where the inequality on the left is imposed because a thin sheet of material
has to be at least one atomic layer thick. 
The total number of carrier charges per atom is defined to be $n_0$.
The packing fraction is always less than unity, $\nu_0<1$.
Thus, the condition $N<1$ states that a one-atom-layer thick material
behaves like a two-dimensional sheet unless the packing fraction is very low.

Using the number density for a two-dimensional sheet from Eq.~(\ref{ne-lim})
in conjunction with Eq.~(\ref{lame=pl}) and $\omega_p^2=n_f e^2/m$
we identify
\begin{equation}
\zeta_p = \frac{e^2}{m} \frac{n_\text{tot}}{A}.
\end{equation}
In this manner $\zeta_p$ is interpreted as a plasma frequency of a 
two-dimensional plasma sheet \cite{Fetter:1973slg,Barton:2005eps}.


\section{Casimir interaction energy between two
 semitransparent $\delta$-function plates}
\label{Cas-tp}

In this section
we calculate the Casimir interaction energy between two parallel 
semitransparent $\delta$-function plates located at $z=a_1$ and $z=a_2$,
with $a=a_2-a_1>0$.
The Casimir interaction energy between two disjoint planar objects
may be calculated using the multiple scattering formalism
\begin{equation}
\frac{E_{12}}{A} = \frac{1}{2} \int_{-\infty}^\infty \frac{d\zeta}{2\pi}
\int \frac{d^2k_\perp}{(2\pi)^2} \text{Tr} \ln \Big[ {\bf 1}
- {\bm \gamma}_1 \cdot {\bf V}_1 \cdot {\bm \gamma}_2 \cdot {\bf V}_2 \Big],
\label{E12-mul}
\end{equation}
where the objects in Eq.~(\ref{E12-mul}) are described by the potentials
representing the dielectric response functions,
${\bf V}_i(z) = {\bm \varepsilon}_i(z) - {\bf 1}$, which 
for $\delta$-function plates are described 
by Eqs.~(\ref{epmu-def-e}) and (\ref{mat-tp-e}) with $\lambda_{ei}^{||}=0$,
with subscripts $i=1,2,$ standing for the respective plates.
See Fig.~\ref{c-d-p-fig}. The trace, $\text{Tr}$, in Eq.~(\ref{E12-mul})
is over both the space coordinate and dyadic index.
For the case under consideration the trace on the coordinate index can
be moved inside the logarithm.
Further, the $\delta$-functions in the potentials from Eq.~(\ref{epmu-def-e})
allow the integrals in the trace over the space coordinate to be performed
and yields
\begin{equation}
\frac{E_{12}^\text{$\delta$-plate}}{A}
= \frac{1}{2} \int_{-\infty}^\infty \frac{d\zeta}{2\pi}
\int \frac{d^2k_\perp}{(2\pi)^2} \text{tr} \ln \Big[ {\bf 1}
- {\bm \gamma}_1(a_2,a_1) \cdot {\bm \lambda}_1 
\cdot {\bm \gamma}_2(a_1,a_2) \cdot {\bm \lambda}_2 \Big],
\label{E12=a2a1}
\end{equation}
in which the trace, $\text{tr}$, is only on the dyadic index.
We point out that semitransparent $\delta$-function plates being considered
are necessarily anisotropic and 
${\bm\lambda}_{ei} = \lambda_{ei}^\perp(i\zeta) \,\text{diag}(1,1,0)$.
Evaluation of ${\bm \gamma}_1(a_2,a_1)$ and ${\bm \gamma}_2(a_1,a_2)$,
see Appendix~\ref{G1-eva-sec} for details,
lets us express Eq.~(\ref{E12=a2a1}) in the form
\begin{equation}
\frac{E_{12}^\text{$\delta$-plate}}{A}
= \frac{1}{2} \int_{-\infty}^\infty \frac{d\zeta}{2\pi}
\int \frac{d^2k_\perp}{(2\pi)^2} \Bigg\{ \ln \Big[ 1
- r_{e1}^H r_{e2}^H\, e^{-2\kappa a}\Big] + \ln \Big[ 1
- r_{e1}^E r_{e2}^E \,e^{-2\kappa a}\Big] \Bigg\},
\label{E12tp-Cas}
\end{equation}
in terms of the reflection coefficients for the separate modes due to
a semitransparent $\delta$-function plate in Eq.~(\ref{rHeg}).
Equation~(\ref{E12tp-Cas}) could have immediately been written down 
once the reflection coefficients were known, because using 
multiple reflections the Casimir energy is determined
by the reflection coefficients geometrically.

For a perfect electrically conducting $\delta$-function plate we take the limit
($\lambda_{ei}^\perp\to\infty$), for which we have
$r_{ei}^H\to 1$, $r_{ei}^E\to -1$.
Using this in Eq.~(\ref{E12tp-Cas}) we have
the interaction energy between two perfect electrically conducting 
$\delta$-function plates,
\begin{equation}
\frac{E_{12}^\text{Cas}}{A} 
= \frac{1}{2} \int_{-\infty}^\infty \frac{d\zeta}{2\pi}
\int\frac{d^2k}{(2\pi)^2}\, 2\,\ln \Big[1  - e^{-2\kappa a}\Big]
=-\frac{\pi^2}{720\,a^3},
\label{E12=ai-lif-einf}%
\end{equation}%
which is exactly the Casimir energy between two 
perfect electrically conducting plates 
\cite{Casimir:1948dh}. This conclusion is in agreement with the result 
in Ref.~\cite{Bordag:2004smr}.

\subsection{Thin-plate limit of 
Casimir interaction energy for two anisotropic slabs}
\label{s-lif-slab}

In Sec.~\ref{physi-dp} we 
showed that the reflection coefficient of a thick dielectric slab 
reduces to that of a semitransparent $\delta$-function plate
in the thin-plate limit of Eq.~(\ref{dtp-con}).
Since the Casimir energy involves integrating over all frequencies
it is of interest to see how the conditions in Eq.~(\ref{dtp-con})
translate for the case of Casimir energies. Specifically, we ask
under what conditions does the Casimir interaction energy of
two anisotropic slabs reduce to that of two semitransparent 
$\delta$-function plates?
To this end we follow the derivation of Lifshitz energy 
between two anisotropic slabs in the spirit of Ref.~\cite{Prachi:2011ec},
with details provided in Appendix~\ref{lifE-aniso-sec}.

\begin{figure}
\subfigure[Semitransparent $\delta$-function plates.]
{\includegraphics[width=5cm]{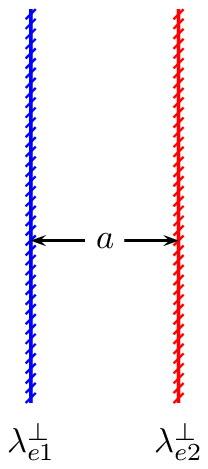}
\label{c-d-p-fig} }
\hspace{15mm}
\subfigure[Dielectric slabs.]
{\includegraphics[width=5cm]{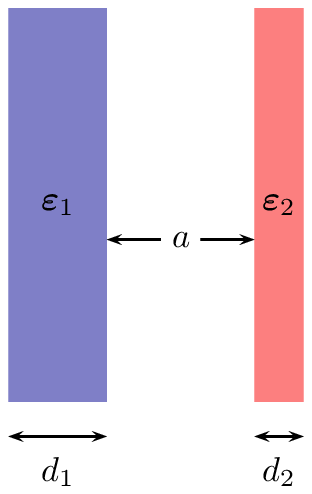}
\label{slabs-fig} }
\caption{ \subref{c-d-p-fig}
Parallel semitransparent $\delta$-function plates separated by distance $a$.
\subref{slabs-fig} Parallel anisotropic dielectric slabs of thicknesses $d_i$
separated by distance $a$.  }
\end{figure}

The interaction energy between two anisotropic slabs is given by
Eq.~(\ref{E12-mul})
when the slabs are described by the potentials
\begin{equation}
{\bf V}_i(z) = ({\bm\varepsilon}_i- {\bf 1}) 
\big[ \theta(z-a_i)-\theta(z-b_i) \big], \qquad i=1,2,
\label{Vi-slab}
\end{equation}
where $b_i-a_i=d_i$ are the thicknesses of the slabs, and $a_2-b_1=a$ is 
the distance between the slabs. See Fig.~\ref{slabs-fig}.
We shall consider an anisotropic dielectric tensor of the form 
${\bm\varepsilon}_i=\text{diag}
(\varepsilon_i^\perp, \varepsilon_i^\perp, \varepsilon_i^{||})$.
The interaction energy of two such slabs is given by
Eq.~(\ref{E12tp-Cas}) with the following replacements in the reflection 
coefficients: $r_{ei}^H \to r_\text{thick}^{H,i}$, and
$r_{ei}^E \to r_\text{thick}^{E,i}$,
see Appendix~\ref{lifE-aniso-sec} for details,
\begin{equation}
\frac{E_{12}^\text{thick}}{A} 
= \frac{1}{2} \int_{-\infty}^\infty\frac{d\zeta}{2\pi}
\int\frac{d^2k}{(2\pi)^2}\, \Bigg\{ \ln \Big[1 
- r_\text{thick}^{H,1} r_\text{thick}^{H,2} \, e^{-2\kappa a}\Big]
+\ln \Big[1 - r_\text{thick}^{E,1} r_\text{thick}^{E,2} \, e^{-2\kappa a}\Big]
\Bigg\},
\label{E12=aiDi-lif}
\end{equation}
which again could have immediately been written down once the 
reflection coefficients were obtained.

\begin{center} \begin{figure}
\includegraphics[width=7cm]{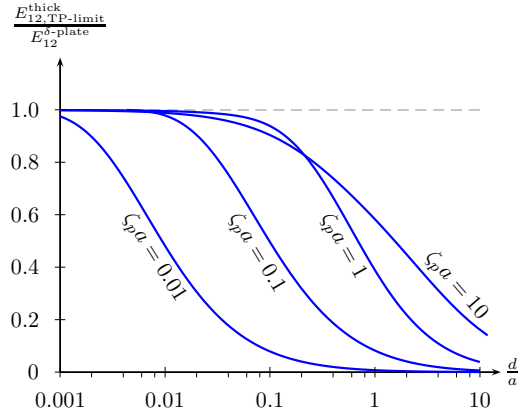}
\caption{Plotted on the ordinate is the ratio of the thin-plate limit
of the interaction energy of two anisotropic dielectric slabs
in Eq.~(\ref{E12=aiDi-lif-cutoff})
to the interaction energy of two $\delta$-function plates      
in Eq.~(\ref{E12tp-Cas}). The fractional error is plotted with 
respect to $d/a$ on a logarithmic scale for different values of $\zeta_pa$.
The ratio approaches unity in the thin-plate limit,
$d/a\ll \zeta_pa\ll a/d$. }
\label{errorintp}
\end{figure} \end{center}

We have earlier shown that when the conditions for the thin-plate limit
in Eq.~(\ref{dtp-con}) are met
the reflection coefficients of a thick anisotropic slab transforms
into the corresponding reflection coefficients
for a $\delta$-function plate, see Eqs.~(\ref{thick-rs}).
Since the thin-plate limit of Eq.~(\ref{dtp-con}) puts bounds on frequencies
and wavenumbers, the thin-plate limit of the Casimir energy between
two anisotropic slabs is estimated by introducing cutoffs in the integrals
of Eq.~(\ref{E12=aiDi-lif}),
\begin{equation}
\frac{E_{12,\text{TP-limit}}^\text{thick}}{A} 
= \frac{1}{2} \int_{-\sqrt{\frac{\zeta_p}{d}}}^{\sqrt{\frac{\zeta_p}{d}}}
\frac{d\zeta}{2\pi}
\int_{-\sqrt{\frac{\zeta_p}{d}}}^{\sqrt{\frac{\zeta_p}{d}}}
\frac{d^2k}{(2\pi)^2}\, \Bigg\{ \quad \cdots \quad \Bigg\},
\label{E12=aiDi-lif-cutoff}
\end{equation}
where the reflection coefficients inside the curly brackets are obtained
using the thin-plate limit of Eq.~(\ref{dtp-con}) as in 
Eqs.~(\ref{thick-rs}).
After scaling the integral variables with the 
distance between the plates we observe that 
Eq.~(\ref{E12=aiDi-lif-cutoff}) is a good approximation of 
the interaction energy between two $\delta$-plates of Eq.~(\ref{E12tp-Cas})
in the parameter regime
\begin{equation}
\frac{d}{a} \ll \zeta_p a \ll \frac{1}{d/a},
\label{tp-cond-da-la}
\end{equation}
which is obtained by rearranging Eq.~(\ref{dtp-con})
after recognizing that typical values of frequencies contributing 
to the integral are of the order $\zeta \sim 1/a$.
This has been illustrated in Fig.~\ref{errorintp} where we plot the
ratio of this thin-plate limit of the interaction energy of two anisotropic
dielectric slabs in Eq.~(\ref{E12=aiDi-lif-cutoff})
to the interaction energy of two $\delta$-function plates 
in Eq.~(\ref{E12tp-Cas}).


\section{Casimir-Polder interaction energy between an atom and
a $\delta$-function plate}
\label{CP-ThickThin}

In this section we calculate the Casimir-Polder interaction energy
between an atom and a semitransparent $\delta$-function plate.
We demonstrate that the same result is obtained in the thin-plate limit
of Eq.~(\ref{tp-cond-da-la}) for the Casimir-Polder interaction energy
between an atom and a dielectric slab.
For a perfect electrically conducting $\delta$-function plate we show that this
energy is exactly equal to the corresponding energy
between an atom and a perfect electrically conducting plate.

\subsection{Atom in front of a $\delta$-function plate}
\label{CP-thin}

We consider an atom with anisotropic electric dipole polarizability
${\bm\alpha} =\text{diag}(\alpha^\perp,
\alpha^\perp, \alpha^{||})$
in front of a semitransparent $\delta$-function plate described
by Eqs.~(\ref{epmu-def-e}) and (\ref{mat-tp-e}) with $\lambda_{ei}^{||}=0$.
The potential for an atom when one neglects quadruple and higher moments is 
${\bf V}({\bf x}) = 4\pi \,{\bm\alpha}(i\zeta) 
\,\delta^{(3)}({\bf x}-{\bf x}_0)$,
where ${\bf x}_0$ is the position of the atom and ${\bm\alpha}(i\zeta)$
is the atomic dipole polarizability of the atom.
An atom in this model is described by at least two parameters,
one corresponding to the static polarizability $|{\bm\alpha}(0)|$,
and the other corresponding to the resonant frequency $\omega_i$ of the atom. 
The weak approximation of the interaction energy in Eq.~(\ref{E12-mul}),
valid for separation distances $r \gg |{\bm\alpha}(0)|^{1/3}$, consists of
retaining only the leading term of the logarithm after expansion.
This approximation is valid for $|{\bm\alpha}(0)|^{1/3}$ small compared to
separation distances. The unretarded (van der Waals-London) regime 
($|{\bm\alpha}(0)|^{1/3} < r\ll c/\omega_i$)
is a short-range approximation, and the retarded (Casimir-Polder) regime 
($|{\bm\alpha}(0)|^{1/3} <c/\omega_i \ll r$)
is the corresponding long-range approximation, where short and long is in 
relation to the characteristic length associated with resonant frequency.
Note that both the van der Waals-London interaction energy and the
Casimir-Polder interaction energy are intrinsically weak.

The expression for Casimir-Polder energy
is obtained by retaining the leading term in the logarithm
and replacing ${\bm\gamma}_2$ with the free Green's dyadic ${\bm\gamma}_0$
in Eq.~(\ref{E12-mul}), which for planar geometries is given by the formula
\begin{equation}
E_{12}^\text{CP} =- 2\pi \int_{-\infty}^\infty \frac{d\zeta}{2\pi}
\int \frac{d^2k_\perp}{(2\pi)^2} \int_{-\infty}^\infty dz
\,\text{tr}\Big[ {\bm\alpha} \cdot
{\bm \gamma}_1(a,z) \cdot {\bf V}_1(z) \cdot {\bm \gamma}_0(z-a) \Big],
\label{CP-pl-gen}
\end{equation}
where trace, $\text{tr}$, is over the dyadic index and $a$
is the distance between the atom, and the $\delta$-function plate is
described by ${\bf V}_1(z)={\bm\lambda}_e\, \delta(z)$,
with ${\bm\lambda}_e=\lambda_e^\perp\,\text{diag}(1,1,0)$.
\begin{figure}
\subfigure[Semitransparent $\delta$-function plate]
{\includegraphics[width=48mm]{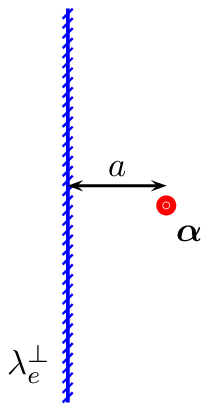}
\label{atom-tp-fig}}
\hspace{20mm}
\subfigure[Dielectric slab]
{\includegraphics[width=48mm]{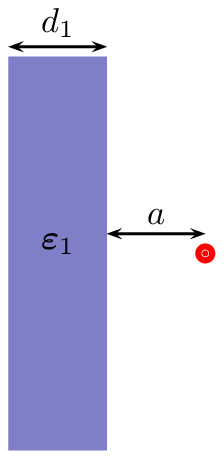}
\label{atom-slab}}
\caption{Anisotropic atom in front of 
\subref{atom-tp-fig} a semitransparent $\delta$-function plate,
versus \subref{atom-slab} an anisotropic dielectric slab.}
\end{figure}
Performing the $z$-integral yields
\begin{equation}
E_{12}^\text{CP} =- 2\pi \int_{-\infty}^\infty \frac{d\zeta}{2\pi}
\int \frac{d^2k_\perp}{(2\pi)^2} \,\text{tr}\Big[ {\bm\alpha}
\cdot {\bm \gamma}_1(a,0) \cdot {\bm \lambda}_e \cdot {\bm \gamma}_0(-a) \Big].
\label{12cp-letp}
\end{equation}
Using Eqs.~(\ref{evGL}) to evaluate ${\bm \gamma}_1(a,0)$,
and using the expression for the free Green's dyadic in the form
\begin{equation}
{\bm\gamma}_0(z) = \left[ \begin{array}{ccc}
-\kappa^2 & 0 & -ik_\perp\kappa\eta(z) \\
0 & -\zeta^2 & 0 \\ -ik_\perp\kappa\eta(z) & 0 & k_\perp^2 
\end{array} \right] \frac{1}{2\kappa} \,e^{-\kappa |z|},
\label{freeG0}
\end{equation}
with $\eta(z)$ defined in Eq.~(\ref{eta-fun}),
into Eq.~(\ref{12cp-letp}), one calculates the Casimir-Polder energy between
an atom and a $\delta$-function plate to be
\begin{equation}
E_{12}^\text{CP} =- 2\pi \int_{-\infty}^\infty \frac{d\zeta}{2\pi}
\int \frac{d^2k_\perp}{(2\pi)^2} \,\frac{e^{-2\kappa a}}{2\kappa}
\Big[ \alpha^\perp (\kappa^2 r_e^H -\zeta^2 r_e^E)
+ \alpha^{||} k_\perp^2 \,r_e^H \Big].
\label{CP12-tp}
\end{equation} 
The above expression for atoms with isotropic polarizabilities
interacting with plates having finite thickness was probably first
reported in Ref.~\cite{Zhou:1995wmw}. In the long distance limit
the Casimir-Polder energy gets contributions from the polarizabilities for
very low frequencies. Thus, we replace
the atomic polarizabilities in Eq.~(\ref{CP12-tp})
by the static polarizabilities ${\bm\alpha}(0)$
though the reflection coefficients are $\zeta$ dependent.
The Casimir-Polder energy between an atom and a perfect electrically
conducting $\delta$-function plate is obtained from Eq.~(\ref{CP12-tp})
by setting $r_e^H=1$ and $r_e^E=-1$, in which case the integrals
in Eq.~(\ref{CP12-tp}) can be completed to yield
\begin{equation}
E_{12}^\text{CP} = - \frac{\text{tr}({\bm\alpha})}{8\pi a^4},
\label{CP12-p}
\end{equation}
which is exactly the Casimir-Polder energy between an atom
and a perfect electrically conducting plate.

\subsection{Thin-plate limit of Casimir-Polder energy for an 
atom in front of an anisotropic dielectric slab} \label{CP=Thick}

Let us next consider an atom in front of an anisotropic dielectric slab
of thickness $d$ described by the potential 
${\bf V}_1(z) = ({\bm\varepsilon}-{\bf 1})[\theta(z+d) -\theta(z)]$,
with ${\bm\varepsilon} = 
\text{diag} (\varepsilon^\perp, \varepsilon^\perp, \varepsilon^{||})$.
See Fig.~\ref{atom-slab}. Using this in Eq.~(\ref{CP-pl-gen}) the 
Casimir-Polder energy for an atom in front of a dielectric slab
is given by
\begin{equation}
E_{12}^\text{CP} =- 2\pi \int_{-\infty}^\infty \frac{d\zeta}{2\pi}
\int \frac{d^2k_\perp}{(2\pi)^2} \int_{-d}^0 dz
\,\text{tr}\Big[ {\bm\alpha} \cdot {\bm \gamma}_1(a,z) \cdot 
({\bm \varepsilon} -{\bf 1}) \cdot {\bm \gamma}_0(z-a) \Big].
\label{12cp-lethick}
\end{equation}
Using Eq.~(\ref{G1ep1}) to evaluate 
${\bm \gamma}_1(a,z) \cdot ({\bm \varepsilon} -{\bf 1})$,
and using Eq.~(\ref{freeG0}) to evaluate the free Green's dyadic,
in Eq.~(\ref{12cp-lethick}) we obtain the Casimir-Polder energy
for an atom in front of an anisotropic dielectric slab
expressed in the form of Eq.~(\ref{CP12-tp}) with the replacements:
$r_e^{H,E} \to r_\text{thick}^{H,E}$,
in terms of the reflection coefficients of the anisotropic slab.
Using the limiting behavior of the reflection coefficients in
Eq.~(\ref{thick-rs}) valid when the conditions in Eq.~(\ref{dtp-con})
are met we then immediately obtain the thin-plate limit to match 
the expression in Eq.~(\ref{CP12-tp}) for the Casimir-Polder energy for
an atom interacting with a $\delta$-function plate. As noted earlier this 
introduces a cutoff on the integrals and thus is valid in the
regime of the thin-plate limit given by Eq.~(\ref{tp-cond-da-la}).


\subsection{Bordag's ``thick'' and ``thin'' boundary conditions}

Bordag \cite{Bordag:2004smr} claims there are two types of perfect 
electrically conducting boundary conditions. In both cases the 
tangential components of the electric field must vanish on the surface $S$,
\begin{equation}
\mathbf{n\times E}\Big|_S=0,\label{usual}
\end{equation}
where $\mathbf{n}$ is the normal vector to the surface.
However, he states that for a ``thick'' conductor, there is an additional
condition on the normal component of the electric field at the surface:
\begin{equation}
\frac\partial{\partial n} E_n\bigg|_S=0,\label{usualbc}
\end{equation}
while for a ``thin'' conductor there is no further constraint on $E_n$.
This distinction is difficult to understand, because Maxwell's equations
should have essentially unique solutions given the boundary condition (\ref{usual})
on a closed surface \cite{Milton:2006vm}.
How indeed is it possible to impose an additional condition on the electric
field?  In fact, Gauss' law in the form ($\varepsilon=1$, Heaviside-Lorentz
units)
\begin{equation}
\bm{\nabla}\cdot\mathbf{E}=\rho,
\end{equation}
in terms of the charge density $\rho$, says, very close to a conducting
surface $S$
\begin{equation}
\frac\partial{\partial n}E_n(\mathbf{r})\bigg|_{\mathbf{r}\to S}=
\delta(\mathbf{r\cdot n})\sigma(\mathbf{r_\perp}),\label{nc}
\end{equation}
which implies the usual statement
\begin{equation}
E_n\Big|_S=\sigma
\end{equation}
on the surface, in terms of the surface charge density $\sigma$.  
The latter for
a conductor is not a specifiable quantity, but must be determined by solving
for the electric field configuration.  The {\it limit\/} of the 
condition (\ref{nc}) as the surface is approached is, in fact, just 
Eq.~(\ref{usual}).

Bordag computes photon propagators in terms of vector potentials in both of
these scenarios.  He shows that if these propagators are used to compute
Casimir forces between conducting bodies, the same results emerge independent
of which propagator is used.  However, when applied to the Casimir-Polder
interaction between a polarizable atom and a conducting plate, the two
propagators give different results.  The ``thick'' propagator gives the
conventional result in Eq.~(\ref{CP12-p}). Using the ``thin'' propagator,
he obtains, for isotropic polarizability, a result which is smaller by a 
factor of 13/15, which is within  the measurement
uncertainty of the experiment \cite{Sukenik:1993fp}.

On the other hand, the Casimir-Polder force can be computed from the electric
Green's dyadic \cite{Schwinger:1977pa}:
\begin{equation}
\bm{\Gamma}(\mathbf{r},t;\mathbf{r'},t')=i\langle \mathbf{E(r},t) \mathbf{E(r'},t')\rangle,\label{ge}
\end{equation}
which is a gauge-invariant quantity.  In terms of the frequency Fourier transform of this,
we can write the Casimir-Polder energy as
\begin{equation}
U_{\rm CP}=\frac{i}2\int\frac{d\omega}{2\pi}4\pi \mbox{tr}\, 
\bm{\alpha}(\omega) \cdot \bm{\Gamma}(\mathbf{R,R},\omega),
\end{equation}
where $\mathbf{R}$ denotes the position of the atom.
In Ref.~\cite{Schwinger:1977pa} the isotropic version of 
Eq.~(\ref{CP12-p}) is obtained,
with no ambiguity in electromagnetic boundary conditions.
However, in the Appendix of that paper,
a propagator in terms of the potentials was also defined:
\begin{equation}
A_\mu(x)=\int d^4x'D_{\mu\nu}(x,x')J^\nu(x')+\partial_\mu \lambda,
\end{equation}
where $J^\nu$ is the (conserved) electric current density, and $\lambda$
is arbitrary gauge function.  Maxwell's equations read
\begin{equation}
L^{\mu\nu}A_\nu=J^\mu,
\end{equation}
where the differential operator is in vacuum ($\varepsilon=\mu=1$)
\begin{equation}
L^{\mu\nu}=\partial^\mu\partial^\nu-g^{\mu\nu}\partial^2,\quad \partial^2=\nabla^2-\partial_0^2.
\end{equation}
The corresponding propagator must satisfy
\begin{equation}
L^{\mu\alpha}L^{\prime\nu\beta}D_{\alpha\beta}=L^{\mu\nu}\delta(x-x').
\label{propeqn}
\end{equation}

Reference~\cite{Bordag:2004smr} 
 presents two propagators, satisfying the thick and thin
boundary conditions, on a conducting surface at $z=0$.  
We will write these in the form
\begin{equation}
D_{\mu\nu}(x,x')=D_{\mu\nu}^{(0)}(x-x')+\bar D_{\mu\nu}(x,x'),
\end{equation}
where the free propagator has the expected form
\begin{equation}
D^{(0)}_{\mu\nu}(x-x')=g_{\mu\nu}\int\frac{d\omega}{2\pi}
\frac{(d^3\mathbf{k})}{(2\pi)^3}e^{-i\omega(t-t')}e^{i\mathbf{k\cdot(r-r')}}
\frac1{k^2},
\end{equation}
in the metric $g_{\mu\nu}=\mbox{diag}(-1,1,1,1)$, $k^2=-(k^0)^2+
\mathbf{k}^2$, $k^0=\omega$.  
We have further noted that any tensor structure in the 
propagator proportional to $\partial_\mu$ or $\partial'_{\nu}$ will
vanish when acted upon by $L^{\mu\alpha}$, $L^{\prime\nu\beta}$.  It is easy
to confirm that $D^{(0)}_{\mu\nu}$ satisfies Eq.~(\ref{propeqn}).  So 
$\bar D_{\mu\nu}$ must satisfy the corresponding homogeneous equation.

Let us first check this for the conventional ``thick'' propagator,
which Bordag writes in the form
\begin{equation}
\bar D_{\mu\nu}(x,x')=
\int\frac{d\omega}{2\pi}
\frac{(d^2\mathbf{k}_\perp)}{(2\pi)^2}\frac{dk_3}{2\pi}
e^{-i\omega(t-t')}e^{i\mathbf{k_\perp\cdot(r-r')_\perp}}e^{ik_3(|z|+|z'|)}
\frac{\bar g_{\mu\nu}}{k^2},\label{corprop}
\end{equation}  
where
\begin{equation}
\bar g_{\mu\nu}=\mbox{diag}(-1,1,1,-1).
\end{equation}
This is constructed so that the $D_{33}$ component vanishes on the plane
$z=0$.  
Now a straightforward calculation shows that
\begin{equation}
L^{\mu\alpha}L^{\prime\nu\beta}\bar D_{\alpha\beta}=\int\frac{d\omega}{2\pi}
\frac{(d^2\mathbf{k}_\perp)}{(2\pi)^2}\frac{dk_3}{2\pi}
e^{-i\omega(t-t')}e^{i\mathbf{k_\perp\cdot(r-r')_\perp}}e^{ik_3(|z|+|z'|)}
M^{\mu\nu}(k),\label{rightside}
\end{equation}
where
\begin{equation}
M^{\mu\nu}(k)=\left(\begin{array}{cccc}
-\mathbf{k}^2&-k^0k_1&-k^0k_2&k^0k_3\\
-k^0k_1&k_1^2+k_3^2-(k^0)^2&-k_1k_2&k_1k_3\\
-k^0k_2&-k_1k_2&k_1^2+k_3^2-(k^0)^2&k_2k_3\\
-k^0k_3&-k_1k_3&-k_2k_3&(k^0)^2-\mathbf{k}^2_\perp
\end{array}\right).
\end{equation}
Notice that the $k^2$ in the denominator has cancelled out, and that therefore
$k_3$ only occurs in the numerator, and consequently when  the integral
over $k_3$ is carried out, we obtain either $\delta(|z|+|z'|)$ or derivatives
thereof.  Since $|z|+|z'|$ is never zero unless both points lie on the surface,
we obtain the required zero value.

The ``thin'' propagator is given by Bordag in terms of specific polarization
vectors. We content ourselves with writing the appropriate metric tensor
in Eq.~(\ref{corprop}),
which follows from those, in terms of  $\kappa^2=\mathbf{k}_\perp^2-(k^0)^2$:
\begin{equation}
\bar g_{\mu\nu}=\sum_{s=1,2}E^s_\mu E^s_\nu
=-\frac1{\kappa^2} \left(\begin{array}{cccc}
\mathbf{k}_\perp^2&-k^0k_1&-k^0k_2&0\\
-k^0k_1&(k^0)^2-k_2^2&k_1k_2&0\\
-k^0k_2&k_1k_2&(k^0)^2-k_1^2&0\\
0&0&0&0\end{array}\right),\label{thinmetric}
\end{equation}
where $E^s_\mu$ is defined in Ref.~\cite{Bordag:2004smr}. 
Then inserting this into the Maxwell equation we obtain a result in the
form of Eq.~(\ref{rightside}) with the tensor
\begin{equation}
\frac{1}{k^2}M^{\mu\nu}(k)=-\frac{1}{\kappa^2}\left(\begin{array}{cccc}
\mathbf{k}_\perp^2&k^0k_1&k^0k_2&0\\
k^0k_1&(k^0)^2-k_2^2&k_1k_2&0\\
k^0k_2&k_1k_2&(k^0)^2-k_1^2&0\\
0&0&0&0\end{array}\right),\label{mthin}
\end{equation}
which differs simply by some signs from Eq.~(\ref{thinmetric}).
Again, this may be argued to be zero because $k_3$ only occurs in the
numerator, so the integration over $k_3$ vanishes.

So the two potential propagators satisfy the same Maxwell equation,
and satisfy, we believe, the same physical boundary conditions, so
they should be gauge-equivalent.  We verify this by showing that they
both correspond to the same electric Green's dyadic, that given in 
Ref.~\cite{Schwinger:1977pa}.  
In that reference, because we are dealing
with a cylindrically symmetric (isotropic) geometry, without loss of
generality we took $\mathbf{k}_\perp=(k_1,0)$.  Then from the usual
construction of the field strength tensor,
\begin{equation}
F^{\mu\nu}=\partial^\mu A^\nu-\partial^\nu A^\mu,
\end{equation}
from which $\mathbf{E}_i=F^{0i}$, and the electric Green's dyadic
of Eq.~(\ref{ge}) can be immediately obtained from the propagators
given above. In fact, using either propagator we find
\begin{equation}
\bm{\Gamma}(\mathbf{r},t;\mathbf{r'},t')=
\bm{\Gamma}^{(0)}(\mathbf{r-r'},t-t')+
\bm{\bar\Gamma}(\mathbf{r},t;\mathbf{r'},t')
\end{equation}
where
\begin{equation}
\bm{\bar\Gamma}(\mathbf{r},t;\mathbf{r'},t')=
\int
\frac{d\omega}{2\pi}\frac{(d^2\mathbf{k_\perp})}{(2\pi)^2}e^{-i\omega(t-t')}
e^{i\mathbf{k_\perp\cdot(r-r')_\perp}}\bar{\bm\gamma}(z,z'),
\end{equation}
with
\begin{equation}
\bar{\bm\gamma}(z,z')=\left(\begin{array}{ccc}
-\kappa^2&0&i\kappa k_\perp\\
0&\omega^2&0\\
-i\kappa k_\perp&0&-k_\perp^2
\end{array}\right)\frac1{2\kappa}e^{-\kappa(|z|+|z'|)},
\end{equation}
which is exactly the Green's dyadic given in Ref.~\cite{Schwinger:1977pa}.
So the two propagators correspond to precisely the same physical
situation, and are just expressed in different gauges.  The apparently
singular $1/\kappa^2$ term (singular in Minkowski space) 
in Eq.~(\ref{mthin}) has no untoward
consequences, but is necessary to establish this coincidence.  Therefore,
we conclude that the usual Casimir-Polder force between a conducting plate
and a polarizable atom is correct, irrespective of whether the plate is
``thick'' or ``thin.''


\section{Conclusion and outlook}

We have considered an infinitesimally thin material
whose optical properties are given in terms of a $\delta$-function.
By integrating across the plate
we have derived the boundary conditions on such a $\delta$-function plate
in Eqs.~(\ref{TM-bc}) and (\ref{TE-bc}).
The Green's dyadic for a $\delta$-function plate,
which completely determines its optical properties, has been derived.
The optical properties of a $\delta$-function plate are shown to necessarily
be anisotropic in that they only depend on the transverse properties 
of the plate. Our work generalizes Barton's insightful 
work on the subject in Refs.~\cite{Barton:2005eps,Barton:2005fst}
for magnetic materials.
We have shown that Bordag's ``thick'' and ``thin''
boundary conditions on a perfect conductor are identical to Barton's 
boundary conditions, which resolves the controversy on the uniqueness
of the boundary conditions for a perfect conductor.

To understand the physical content of
our results for a $\delta$-function plate we
defined the thin-plate limit, given by Eq.~(\ref{dtp-con}) for 
reflection coefficients and by Eq.~(\ref{tp-cond-da-la}) for Casimir energies,
and clarified how a $\delta$-function plate is physically realized as
an infinitesimally thin plate. We have shown that
in the thin-plate limit of Eq.~(\ref{dtp-con}) the reflection
coefficients of a dielectric slab reduce to those of a $\delta$-function plate.
The thin-plate limit of the interaction energy between two anisotropic slabs
is shown to be the Casimir energy between two $\delta$-function plates.
Similarly, the thin-plate limit of the interaction energy between an
atom and an anisotropic slab is shown to be the Casimir-Polder energy
between an atom and a $\delta$-function plate. 

The reflection coefficients obtained for a $\delta$-function plate
in Eq.~(\ref{rHeg}) are very general because they are independent of
the dispersion model used to represent
$\lambda_e^\perp(i\zeta)$ and $\lambda_g^\perp(i\zeta)$.
In our discussions, from Sec.~\ref{physi-dp} onwards,
we exclusively used the plasma model for a 2-dimensional metal
to represent $\lambda_e^\perp(i\zeta)$.
Since the thin-plate limit of Eq.~(\ref{dtp-con}) is a constraint on
the frequency response, the definition of thin-plate limit will depend on
the dispersion model used to represent $\lambda_e^\perp(i\zeta)$.
Thus, the thin-plate limit defined in Eq.~(\ref{dtp-con})
is specific for a plasma model and is sufficient as long as the 
relevant frequencies are large compared to relaxation frequency.
The corresponding thin-plate limit when the low frequency limit of
the Drude model represents
$\lambda_e^\perp(i\zeta)$ is being employed by us to model graphene,
which we believe will lead to interesting insights into the
quantum electromagnetic interactions of graphene.

\acknowledgments

We thank Gabriel Barton, Michael Bordag, Iver Brevik, Simen Ellingsen,
and Aram Saharian, for helpful suggestions and comments,
and Elom Abalo and Nima Pourtolami for collaborative assistance.
PP thanks the University of Zaragoza for hospitality and
In\'{e}s Cavero-Pel\'{a}ez and Manuel Asorey for helpful discussions.
PP acknowledges the support of the Julian Schwinger Foundation and
the European Science Foundation during the course of this work.
KVS would like to thank the Homer L. Dodge Department of Physics 
and Astronomy at the University of Oklahoma for hospitality and support,
and K. V. Jupesh and Dipayan Paul for discussions.
MS and KVS were supported by the National Science Foundation with
Grant No.~PHY0902054. KAM and PP
acknowledge support from the National Science Foundation 
with Grant No.~PHY0968492
and the Department of Energy with Grant No.~DE-FG02-04ER41305.

\appendix

\section{Proof for ``averaging prescription''}
\label{avpres-sec}

Let us consider a function $f(x)$ with a step discontinuity
\begin{equation}
f(x) = a_1 \,\theta(-x) + a_2 \,\theta (x),
\end{equation}
written using the $\theta$-function of Eq.~(\ref{th-f-def}),
whose value at $x=0$, $f(0)$, is not well defined.
If we now evaluate the integral, using the standard prescription,
we obtain
\begin{equation}
\int_{-\infty}^\infty dx \,\delta (x) f(x) = f(0),
\label{tesint}
\end{equation}
which in turn now is not well defined.

The $\delta$-function can be interpreted as a limit of a sequence of 
functions 
\begin{equation}
\delta (x) = \lim_{\epsilon\to 0} u_\epsilon (x),
\label{dx=limu}
\end{equation}
each of which satisfies
\begin{equation}
\int_{-\infty}^\infty dx \,u_\epsilon(x) = 1.
\end{equation}
For our purpose, we shall restrict ourselves to functions that are 
symmetric with respect to the origin \cite{Estrada:2007pt}
\begin{equation}
u_\epsilon(-x) = u_\epsilon(x).
\label{uep-sym}
\end{equation}

Using the representation of the $\delta$-function in Eq.~(\ref{dx=limu})
to evaluate the integral in Eq.~(\ref{tesint}) we obtain
\begin{subequations}
\begin{eqnarray}
\int_{-\infty}^\infty dx\,\delta(x) f(x)
&=& a_1 \, \lim_{\epsilon\to 0} \int_{-\infty}^0 dx \,u_\epsilon (x)
 + a_2 \, \lim_{\epsilon\to 0} \int_0^\infty dx\,u_\epsilon (x) \\
&=& \left( \frac{a_1+a_2}{2} \right) \int_{-\infty}^\infty dx\,u_\epsilon (x)
\\ &=& \frac{a_1+a_2}{2}, \label{avf0} 
\end{eqnarray}
\end{subequations}
where we used the symmetry restriction of Eq.~(\ref{uep-sym}) in the
second equality. Using the result in Eq.~(\ref{avf0}) to define the 
value of $f(0)$ in Eq.~(\ref{tesint}) we have
\begin{equation}
f(0) = \frac{a_1+a_2}{2},
\end{equation}
which is the averaging prescription.

We shall point to Ref.~\cite{Estrada:2007pt} 
for discussions and references on this topic.


\section{Evaluation of Green's dyadics for a $\delta$-function plate}
\label{G1-eva-sec}

Evaluation of ${\bm \gamma}_1(a_2,a_1)$ in Eq.~(\ref{E12-mul})
is performed using Eq.~(\ref{Gamma=gE})
in terms of the magnetic Green's function in Eq.~(\ref{gHtp})
with $r_g^H=0$ and the dielectric functions of
the media appearing in Eq.~(\ref{Gamma=gE}) being set to unity because
we are considering $\delta$-function plates in vacuum.
The corresponding electric Green's function is obtained by interchanging
the electric and magnetic properties. Furthermore, these Green's functions
are evaluated on the position of the plates using the averaging prescription 
\cite{CaveroPelaez:2008tj}. In the evaluation of ${\bm \gamma}_1(a_2,a_1)$
we use 
\begin{subequations}
\begin{eqnarray}
g_1^E(z,z^\prime)\big|_{z=a_2,z^\prime=a_1}
&=& \frac{1}{\zeta^2}
\frac{1}{\left[\lambda_{e1}^\perp(i\zeta)+ 
   \frac{2\kappa}{\zeta^2}\right]} \,e^{-\kappa a}, \\
\partial_{z^\prime} g_1^H(z,z^\prime)\big|_{z=a_2,z^\prime=a_1}
&=& \frac{1}{\kappa}
\frac{1}{\left[\lambda_{e1}^\perp(i\zeta)+ 
   \frac{2}{\kappa}\right]} \,e^{-\kappa a}, \\
\partial_z \partial_{z^\prime} g_1^H(z,z^\prime)\big|_{z=a_2,z^\prime=a_1}
&=& - \frac{1}{\left[\lambda_{e1}^\perp(i\zeta)+ 
   \frac{2}{\kappa}\right]} \,e^{-\kappa a},
\end{eqnarray}%
\label{evGL}
\end{subequations}
which using Eq.~(\ref{Gamma=gE}) yields ${\bm \gamma}_1(a_2,a_1)$.


\section{Lifshitz interaction energy for anisotropic dielectric slabs}
\label{lifE-aniso-sec}

For the case of two anisotropic slabs (with isotropy in the plane of
the slabs) the Green's dyadic in Eq.~(\ref{Gamma=gE})
can be decomposed into magnetic and electric modes,
and we can can write the Casimir energy per unit area in the form
\begin{equation}
\frac{E_{12}^\text{thick}}{L_xL_y} 
= \frac{1}{2} \int_{-\infty}^\infty\frac{d\zeta}{2\pi}
\int\frac{d^2k_\perp}{(2\pi)^2}\, \Big\{ \ln \big[1-K^E\big] +
\ln \big[1-K^H\big] \Big\},
\end{equation}
where 
\begin{eqnarray}
K^E &=& (\varepsilon_1^\perp-1)(\varepsilon_2^\perp-1)\zeta^4
\int_{a_1}^{b_1} dz \int_{a_2}^{b_2} dz^\prime\,
g^E_{1}(z^\prime,z) g^E_{2}(z,z^\prime), \\
K^H &=& \int_{a_1}^{b_1} dz \int_{a_2}^{b_2} dz^\prime\,
\,\text{tr} \Big[ {\bm\gamma}^H_{1}(z^\prime,z) \cdot 
({\bm\varepsilon}_1-{\bf 1}) \cdot {\bm\gamma}^H_{2}(z,z^\prime)
\cdot ({\bm\varepsilon}_2-{\bf 1}) \Big].
\end{eqnarray}
The regions for evaluation of $z$ and $z^\prime$ are unambiguously
specified by the integration regions.
Notice that we can omit the $\delta$-function term in Eq.~(\ref{Gamma=gE})
since $z$ and $z^\prime$ are never evaluated at the same point
for disjoint objects.
Thus we can write
\begin{equation}
{\bm\gamma}^H_{1}(z^\prime,z) \cdot ({\bm\varepsilon}_1-{\bf 1})
= \left[ \begin{array}{cc}
\partial_z \partial_{z^\prime} g^H_{1}(z^\prime,z) 
& ik_\perp \partial_{z^\prime} g^H_{1}(z^\prime,z) \\[2mm]
-ik_\perp \partial_{z} g^H_{1}(z^\prime,z) & k_\perp^2 g^H_{1}(z^\prime,z)
\end{array} \right]
\left[ \begin{array}{cc}
(\varepsilon_1^\perp -1)/\varepsilon_1^\perp & 0 \\[2mm]
0 & (\varepsilon_1^{||} -1)/\varepsilon_1^{||}
\end{array} \right],
\label{G1ep1}
\end{equation}
where we have used the fact that the Green's function for the first slab,
$g_1^H(z^\prime,z)$, evaluated for $a_1<z<b_1$ and $a_2<z^\prime<b_2$
corresponds to choosing ${\bm \varepsilon}_1(z) = {\bm \varepsilon}_1$, and
${\bm \varepsilon}_1(z^\prime) = {\bf 1}$. Similarly,
\begin{equation}
{\bm\gamma}^H_{2}(z,z^\prime) \cdot ({\bm\varepsilon}_2-{\bf 1})
= \left[ \begin{array}{cc}
\partial_z \partial_{z^\prime} g^H_{2}(z,z^\prime) 
& ik_\perp \partial_{z} g^H_{2}(z,z^\prime) \\[2mm]
-ik_\perp \partial_{z^\prime} g^H_{2}(z,z^\prime) 
& k_\perp^2 g^H_{1}(z,z^\prime)
\end{array} \right]
\left[ \begin{array}{cc}
(\varepsilon_2^\perp -1)/\varepsilon_2^\perp & 0 \\[2mm]
0 & (\varepsilon_2^{||} -1)/\varepsilon_2^{||}
\end{array} \right],
\end{equation}
observing that the Green's function for the second slab,
$g_2^H(z,z^\prime)$, evaluated for $a_1<z<b_1$ and $a_2<z^\prime<b_2$
corresponds to choosing ${\bm \varepsilon}_2(z) = {\bf 1}$, and
${\bm \varepsilon}_2(z^\prime) = {\bm \varepsilon}_2$.

The magnetic Green's functions for $a_1<z<b_1$ and $a_2<z^\prime<b_2$ 
for the respective slabs are \cite{Prachi:2011ec}
\begin{subequations}
\begin{eqnarray}
g^H_{1}(z^\prime,z) &=& \,
\frac{e^{-\kappa (z^\prime-b_1)} 
\big[ (\bar{\kappa}_1^H+\kappa) e^{-\kappa_1^H(b_1-z)} 
+ (\bar{\kappa}_1^H-\kappa) e^{-\kappa_1^Hd_1} e^{-\kappa_1^H(z-a_1)} \big]}
{\big[(\bar{\kappa}_1^H+\kappa)^2
- (\bar{\kappa}_1^H-\kappa)^2 e^{-2\kappa_1^Hd_1}\big]}, \\
g^H_{2}(z,z^\prime) &=& \,
\frac{ e^{-\kappa (a_2-z)} 
\big[ (\bar{\kappa}_2^H+\kappa) e^{-\kappa_2^H(z^\prime-a_2)} 
+ (\bar{\kappa}_2^H-\kappa) 
e^{-\kappa_2^Hd_2} e^{-\kappa_2^H(b_2-z^\prime)} \big]}
{\big[ (\bar{\kappa}_2^H+\kappa)^2 -
(\bar{\kappa}_2^H-\kappa)^2 e^{-2\kappa_2^Hd_2} \big] },
\end{eqnarray}
\end{subequations}
and the corresponding electric Green's function is obtained by 
swapping the electric and magnetic properties. We have used 
the definitions for $\kappa_i^H$'s and $\kappa_i^E$'s
in Eqs.~(\ref{kaiH}) and (\ref{kaiE}) with ${\bm\mu}={\bf 1}$.
Completing the algebra we have the interaction energy between
parallel anisotropic slabs presented in Eq.~(\ref{E12=aiDi-lif})
and written in terms of the reflection coefficients for a thick 
anisotropic slab introduced in Eqs.~(\ref{thick-rs}).


\bibliography{%
biblio/b20111229-bc-itp,%
}


\end{document}